\documentclass[12pt,preprint]{aastex}

\shorttitle{Confined flare} \shortauthors{Li et al.}

\begin{document}

\title{Two Types of Solar Confined Flares}

\author{Ting Li\altaffilmark{1,2}, Lijuan Liu\altaffilmark{3,4}, Yijun Hou\altaffilmark{1,2} \& Jun Zhang\altaffilmark{1,2}}

\altaffiltext{1}{CAS Key Laboratory of Solar Activity, National
Astronomical Observatories, Chinese Academy of Sciences, Beijing
100101, China; liting@nao.cas.cn} \altaffiltext{2}{School of
Astronomy and Space Science, University of Chinese Academy of
Sciences, Beijing 100049, China} \altaffiltext{3}{School of
Atmospheric Sciences, Sun Yat-sen University, Zhuhai, Guangdong,
519082, China; liulj8@mail.sysu.edu.cn} \altaffiltext{4}{CAS Center
for Excellence in Comparative Planetology, China}

\begin{abstract}

With the aim of understanding the physical mechanisms of confined
flares, we selected 18 confined flares during 2011-2017, and
classified the confined flares into two types based on their
different dynamic properties and magnetic configurations. ``Type I"
of confined flares are characterized by slipping reconnection,
strong shear, and stable filament. ``Type II" flares have nearly no
slipping reconnection, and have a configuration in potential state
after the flare. Filament erupts but is confined by strong strapping
field. ``Type II" flares could be explained by 2D MHD models while
``type I" flares need 3D MHD models. 7 flares of 18 ($\sim$39 \%)
belong to ``type I" and 11 ($\sim$61 \%) are ``type II" confined
flares. The post-flare loops (PFLs) of ``type I" flares have a
stronger non-potentiality, however, the PFLs in ``type II" flares
are weakly sheared. All the ``type I" flares exhibit the ribbon
elongations parallel to the polarity inversion line (PIL) at speeds
of several tens of km s$^{-1}$. For ``type II" flares, only a small
proportion shows the ribbon elongations along the PIL. We suggest
that different magnetic topologies and reconnection scenarios
dictate the distinct properties for the two types of flares.
Slipping magnetic reconnections between multiple magnetic systems
result in ``type I" flares. For ``type II" flares, magnetic
reconnections occur in anti-parallel magnetic fields underlying the
erupting filament. Our study shows that ``type I" flares account for
more than one third of the overall large confined flares, which
should not be neglected in further studies.

\end{abstract}


\section{Introduction}

Solar flares are among the most energetic phenomena in the solar
atmosphere filled with magnetized plasma. They are often associated
with coronal mass ejections (CMEs), which are the dominant
contributor to adverse Space Weather at Earth (e.g., Gosling et al.
1991). It is widely believed that the flares and CMEs are two
manifestations of the same underlying physical processes. The flares
associated with a CME are usually referred to be eruptive events,
while the flares that are not accompanied by a CME are called
confined events (Svestka \& Cliver 1992). Although confined flares
have no influence on our space weather conditions, the better
understanding of confined flares helps to reveal the physical
mechanism of flares and their relationship with CMEs.

The process of eruptive two-ribbon flares is usually described by
the CSHKP model (Carmichael 1964; Sturrock 1966; Hirayama 1974; Kopp
\& Pneuman 1976) and its extension in three dimensions (3D, Aulanier
et al. 2012; Janvier et al. 2014). In the CSHKP model, magnetic
energy is stored in sheared magnetic arcades or a core flux rope
above the polarity inversion line (PIL). Due to the loss of
equilibrium, the core magnetic flux system (sheared or twisted)
starts to move upward and stretches the embedding magnetic fields.
Thus the current sheet is formed underlying the rising core magnetic
flux system, and magnetic reconnection occurring in the current
sheet releases large amounts of energy via a reconfiguration of the
magnetic connectivity. The accelerated particles propagate along the
reconnected field lines towards the denser lower atmosphere and
generate flare loops and ribbons. As the magnetic reconnection goes
on, the reconnection sites are ascending and the flare ribbons are
widely observed to separate from each other perpendicularly to the
PIL. The erupting core magnetic flux system propels plasma into the
interplanetary space and forms a CME.

In the classical two-dimensional (2D) scenario, the preferred sites
for the formation of current sheets are the topological features
such as magnetic null points and separatrix, which highlight
discontinuities in magnetic connectivity (Priest \& Forbes 2002).
Different from the 2D mode, magnetic reconnection in three
dimensions (3D) can also occur at the regions of very strong
magnetic connectivity gradients, i.e., quasi-separatrix layers
(QSLs, D{\'e}moulin et al. 1996; Chandra et al. 2011). The close
correspondence between flare ribbons and the photospheric signature
of QSLs has been shown in many studies (Schmieder et al. 1997;
Masson et al. 2009; Zhao et al. 2014; Savcheva et al. 2015; Yang et
al. 2015), which provides a strong evidence for the QSL
reconnection. When magnetic reconnection occurs along the QSL,
magnetic connectivity is continuously exchanged between neighboring
field lines, and the magnetic field lines are observed to ``slip"
inside the plasma, named slipping reconnection in sub-Alfv\'{e}nic
regime (Priest \& D{\'e}moulin 1995; Aulanier et al. 2006). In
recent years, the standard 3D flare model has been proposed to
interpret the intrinsic 3D nature of eruptive flares (Aulanier et
al. 2012; Janvier et al. 2014). In this model, the magnetic flux
rope is surrounded by the QSLs and the current layer forms along the
QSLs. The magnetic reconnection occurs below the erupting flux rope
when the current layer at the QSL becomes thin enough, and the flare
ribbons coincide with the double J-shaped QSL footprints. Recent
high-quality imaging and spectroscopic observations have revealed
the signatures of QSL reconnections during eruptive flares
(Dud{\'{\i}}k et al. 2014, 2016; Li \& Zhang 2015; Zheng et al.
2016; Gou et al. 2016; Li et al. 2016; Jing et al. 2017). The flare
loops were observed to slip along the developing flare ribbons at
speeds of several tens of km s$^{-1}$ (Dud{\'{\i}}k et al. 2014,
2016). Two flare ribbons exhibited the elongation motions in
opposite directions along the polarity inversion line (Li \& Zhang
2014), and the ribbon substructures were seen to undergo a
quasi-periodic slipping motion along the ribbon (Li \& Zhang 2015).

For confined flares, the key question is the factor determining the
confined character of solar flares. Wang \& Zhang (2007) analyzed
eight X-class flares and found that the confined events occur closer
to the magnetic center and the eruptive events tend to occur close
to the edge of active regions (ARs), implying that the strong
external field overlying the AR core is probably the main reason for
the confinement. Similar results have been found by Baumgartner et
al. (2018) based on a statistical analysis of 44 flares during
2011-2015. Amari et al. (2018) suggested that the role of the
magnetic cage crucially affects the class of eruption$-$either
confined or eruptive. To date, many studies have made the consistent
conclusion that the decay index of the potential strapping field
determines the likelihood of ejective/confined eruptions (Green et
al. 2002; Shen et al. 2011; Yang et al. 2014; Thalmann et al. 2015;
Chen et al. 2015; Li et al. 2018a). Previous studies showed that the
torus instability of a magnetic flux rope occurs when the critical
decay index reaches 1.5 (Bateman 1978; Kliem \& T{\"o}r{\"o}k 2006).
Zuccarello et al. (2015) performed a series of numerical
magnetohydrodynamics (MHD) simulations to flux rope eruptions and
found that the critical decay index for the onset of the torus
instability lies in a range of 1.3-1.5. Another factor determining
whether a flare event is CME-eruptive or not is the non-potentiality
of ARs including the free magnetic energy, relative helicity, and
magnetic twists (Falconer et al. 2002, 2006; Nindos \& Andrews 2004;
Tziotziou et al. 2012). Sun et al. (2015) suggested that AR
eruptiveness is related to relative value of magnetic
non-potentiality over the constraint of background field. However,
in the statistical study of Jing et al. (2018), the unsigned twist
number of magnetic flux rope plays little role in differentiating
between confined and ejective flares and the decay index of the
potential strapping field above the flux rope well discriminates
them.

In recent years, abundant observations of solar flares show that
some flares are not consistent with the classical flare models. This
kind of atypical flares has been investigated by several case
studies. Liu et al. (2014) carried out the first topological study
of an ``unorthodox" X-class confined flare exhibiting a cusp-shaped
structure, and concluded that the QSL reconnections at the T-type
hyperbolic flux tube above the flux rope result in the dynamics of
nested loops within the cusp. In the study of Dalmasse et al.
(2015), the atypical confined flare was caused by multiple and
sequential magnetic reconnections occurring in a complex magnetic
configuration of several QSLs. In the recent paper of Joshi et al.
(2019), they presented another case study of an atypical confined
flare and found that a curved magnetic polarity inversion line of
the AR is a key ingredient for producing the atypical flares.
However, all of these works are case studies and it is not clear
about the physical characteristics and reconnection process of
atypical confined flares due to the lack of statistical studies.

In our study, we carry out a statistical analysis of 18 confined
flares larger than M5.0 class during 2011-2017. Based on the flare
dynamics and extrapolated coronal magnetic fields, we first classify
the confined flares into two types. In ``type I" confined flares,
multiple slipping magnetic reconnections occur in a complex magnetic
configuration along two or more QSLs overlying the core magnetic
structure, and the entire magnetic system involved in the flare
still remains stabilized. ``Type II" has the magnetic configuration
consistent with the classical flare models, but strong strapping
fields are present over the flaring region in the high atmosphere.
The structure of the paper is as follows. In Section 2, we describe
our event sample and the extrapolated method. Section 3 presents the
detailed analysis for three events as typical examples. Finally, in
Section 4 we discuss our results and conclude in Section 5.

\section{Observations and Data Analysis}

In this study, we examined the Geostationary Operational
Environmental Satellite (\emph{GOES}) soft X-ray (SXR) flare
catalog\footnote{\url{ftp://ftp.ngdc.noaa.gov/STP/space-weather/solar-data/solar-features/solar-flares/x-rays/goes/xrs/}}
to search for flare events larger than M5.0 within $45^{\circ}$ from
the disk center in the period from 2011 January to 2017 December.
For each event, the CME
catalog\footnote{\url{https://cdaw.gsfc.nasa.gov/CME\_list/}}
(Gopalswamy et al. 2009) of the Solar and Heliospheric Observatory
(\emph{SOHO})/Large Angle and Spectrometric Coronagraph (LASCO) was
checked to determine whether the flare was confined or not. We
regarded a flare as confined if there is no CME within 60 min of the
flare start time in the quadrant consistent with the flare position.
In addition, we also visually inspected the \emph{Solar Dynamics
Observatory} (\emph{SDO}; Pesnell et al. 2012) and the twin Solar
Terrestrial Relations Observatory (\emph{STEREO}; Kaiser et al.
2008; Howard et al. 2008) observations to further verify the
classification of eruptive and confined flares. A total of 18
confined flares from 12 ARs fulfilled the selection criteria above
(see Table 1).

To analyze the structure and dynamics of each confined flare, we
used the E(UV) observations from the Atmospheric Imaging Assembly
(AIA; Lemen et al. 2012) on board the \emph{SDO}, with a resolution
of $\sim$0$\arcsec$.6 per pixel and a cadence of 12/24 s. Four
channels of AIA 1600, 304, 171, and 131 {\AA} were mainly applied to
classify these confined flares. The topology of the magnetic fields
of the AR is important to interpret the development of the flare.
Thus we performed the nonlinear force-free field (NLFFF)
extrapolations (Wheatland et al. 2000; Wiegelmann 2004) to the
selected events and obtained their 3D source coronal magnetic
fields. The vector magnetogram used as the bottom boundary condition
(z=0) for the extrapolations was an Helioseismic and Magnetic Imager
(HMI; Scherrer et al. 2012) data product called Space-Weather HMI AR
Patches (Bobra et al. 2014). The vector magnetogram was
pre-processed to remove the net force and torque on the photospheric
boundary (Wiegelmann et al. 2006). We also calculated the squashing
factor (Q) and twist number based on the extrapolated 3D magnetic
fields with the code introduced by Liu et al. (2016b). The QSL
distribution can be quantified by the Q factor, which measures the
magnetic field connectivity gradient by tracing field lines
pointwise (Titov et al. 2002).

\section{Results}

We looked through the AIA 1600, 304, 171, and 131 {\AA} movies for
all the 18 confined flares, and found that our events could be
categorized into two groups:

``Type I" flares.-The common characteristic among all the flares in
this group is that the flaring structure is complex with two or more
sets of magnetic systems, and it is not associated with any eruption
of core filament. The post-flare loops (PFLs) are strongly sheared
with respect to the PIL of the AR and they are overlying the
non-eruptive filament.

``Type II" flares.-The flares in this group are associated with the
failed eruption of a core filament, and have the weakly sheared PFLs
underlying the erupting filament. Large-scale strapping loop bundles
overlying the flaring region could be seen in high-temperature
wavelengths (131 and 94 {\AA}) during the development of the flare.
In other words, this group of flares are consistent with the
classical flare models.

According to the classification criteria of confined flares, 7
flares of 18 ($\sim$39 \%) belong to ``type I" and 11 ($\sim$61 \%)
are ``type II" confined flares (see Table 1). Two events of ``type
I" and one event of ``type II" are taken as examples to analyze the
flare dynamics and magnetic topological structures in detail.

\subsection{``Type I": the X1.5-class Flare on 2011 March 09}

One selected event of ``type I" confined flares is the X1.5-class
event occurring in AR 11166 near the solar disk center
(N$08^{\circ}$, W$11^{\circ}$) on 2011 March 09. The GOES soft X-ray
1$-$8 {\AA} flux showed that the X1.5-class flare initiated at 23:13
UT and reached its peak at 23:23 UT (Figure 3(f)). Line-of-sight
(LOS) magnetograms from the HMI on board the \emph{SDO} are used to
investigate the evolution of photospheric magnetic fields before the
flare. The evolution of the AR presented localized magnetic flux
emergence and strong shearing motions as displayed in Figure 1. The
new negative-polarity patch N2 emerged nearby the pre-existing
negative-polarity patch N1 about one day before the flare onset.
Simultaneously, the emerging patch N2 showed a strong shearing
motion towards the northwest, together with a weak shear of
pre-existing patch N1 along the same direction with N2. Along the
shearing direction of patch N2 (white dash-dotted curve ``S1" in
Figure 1(a)), we obtain a stack plot (Figure 1(e)) based on the
12-min LOS magnetograms. The average shearing speed of emerging
patch N2 was about 0.34 km s$^{-1}$, comparable to the statistical
results showing that the maximum shear-flow speeds of the above M1.0
flaring-AR have a peak value of 0.3$-$0.4 km s$^{-1}$ (Park et al.
2018; Hou et al. 2018).

At the flaring region, three filaments F1-F3 existed between the
positive and negative magnetic fluxes (Figures 2(a) and (d)). Their
lengths are about 10-30 Mm, belonging to the category of
mini-filaments (Hermans \& Martin 1986; Hong et al. 2017). During
the flare process, these filaments did not show the rise phase and
were not associated with any failed eruptions. They were stably
present after the flare and seemed not to be affected by the
evolution of the flare (Figure 2(b)). For ``type I" confined flares,
high-temperature flare loops displayed significant dynamic
evolution, and thus AIA 94 and 131 {\AA} observations (about 7 MK
for 94 {\AA} and 11 MK for 131 {\AA}; O'Dwyer et al. 2010) were
analyzed in detail. From about 22:47 UT ($\sim$26 min before the
flare onset), the flare loops started to be illuminated in 131 {\AA}
channel indicating the initiation of magnetic reconnections. We
identified two sets of bright loop bundles labeled L1-L4 in the hot
line (131 {\AA}) at 22:56 UT (Figure 2(i)), displaying two different
magnetic connectivities. These loop bundles can not be clearly
discerned in 94 {\AA} (Figure 2(e)), meaning that their temperature
is around 10 MK. As the flare developed, more flare loops appeared
and delineated ``fan-shaped surfaces" (Figures 2(f) and (j)). At
about 23:20 UT, new loop bundles NL1 and NL2 were formed (Figures
2(g) and (k)), implying the reconfiguration of magnetic fields
caused by magnetic reconnections. Later on, the 94 {\AA}
observations showed the formation of another two loop bundles NL3
and NL4 at 23:25 UT (Figure 2(h)). The PFLs detected in the
low-temperature line (171 {\AA} at $\sim$0.6 MK) were formed
overlying the non-eruptive filaments (Figure 2(l)).

The comparison of the 1600 {\AA} image with HMI LOS magnetogram
showed that the flare consisted of two positive-polarity ribbons R1
and R2 and a semi-circular ribbon R3 (Figures 2(c)-(d)). The
pre-flare evolution of magnetic fields showed that ribbon R3 was
located at the emerging and shearing negative-polarity patch N2
(Figures 1(d) and 2(d)). We overplotted the pre-flare loop bundles
L1-L4 of 131 {\AA} on the 1600 {\AA} image and found that their
footpoints were perfectly co-spatial with the flare ribbons. The
eastern footpoints of loops L1-L2 were located at ribbon R1 and
their western footpoints at ribbon R3. The north ends of loops L3-L4
anchored in ribbon R2 and their south ends at ribbon R3. The
correspondence of the loop footpoints with the flare ribbons implies
that the magnetic reconnections mainly occur in the two sets of
magnetic connectivities outlined by loops L1-L4. We suggest that L2
and L3 are reconnecting, then L1 and L4 move toward the reconnection
region and reconnect subsequently. During the development of the
flare, ribbons R1-R3 did not show any evident separation motion
perpendicular to the PIL (orange line in Figure 2(d)). Ribbons R1
and R3 exhibited the elongation motions at their hook parts along
the magnetic PIL. Ribbon brightening of R1 spread towards the south
and R3 appeared to spread mostly northward at an average speed of 33
km s$^{-1}$ (see Table 1).

The flare loops exhibited an apparent slipping motion towards the
northwest in the impulsive of the flare in 94 {\AA} channel (Figure
3). From about 23:15 UT, the loop bundle connecting ribbons R2 and
R3 was seen to slip. The easternmost loop of the bundle can be
clearly discerned (SL1-SL4 in Figures 3(a)-(d)) and thus was used to
trace the dynamic evolution of the loop bundle. The south footpoints
of the slipping loops SL1-SL4 moved along ribbon R3 to the west, and
the displacement of their north footpoints along ribbon R2 was not
evident compared with that of the south ends. The morphology of the
loops was changed gradually from curved to straight, suggesting that
each time it was not the same structure, but a new flare loop. The
new flare loop became visible due to the slipping magnetic
reconnections, similar to the observations of Li \& Zhang (2014,
2015). Eventually, these successively visible loop structures
delineated a ``fan-shaped surface" (Figure 3(d)). To analyze the
slipping motion, we placed an artificial cut ``S2" along the
slipping direction (white dotted curve in panel (a)). This cut was
used to produce the time-distance plot in AIA 94 {\AA} shown in
Figure 3(e). As shown from the time-distance plot, the slipping
motion exhibited an acceleration process during the flare evolution.
By tracing the easternmost loop in the time-distance plot (dotted
line in panel (e)), we obtained the velocity-time plot of the
slipping motion (black curve in Figure 3(f)). The slippage can be
described by two kinematic phases: a slow slipping phase and a fast
slipping phase. The slipping motion was slow in the early stage and
reached $\sim$11 km s$^{-1}$ at about 23:17 UT. Then the apparent
slipping speed started to increase impulsively and up to $\sim$84 km
s$^{-1}$ at about 23:21 UT. We assumed an error of two pixels
(1$\arcsec$.2) in the height measurement and the uncertainty in the
speed was estimated to be about 8.5 km s$^{-1}$. The speed profile
of the slippage has a similar trend to the GOES SXR 1$-$8 {\AA} flux
variation (blue curve in panel (f)), but with a delay of several
minutes.

We investigated the development of PFLs in the gradual phase of the
flare and estimated the inclination angles of PFLs with respect to
the PIL (Figure 4). The average orientation of the PIL is determined
from the HMI LOS magnetogram prior to the flare onset. The
inclination angle $\theta$ corresponds to the angle between the
tangents of the PFL and PIL at their intersection (panel (b)), which
is consistent with the method of Zhang et al. (2017). The
complementary angle of $\theta$ has been referred to as the shear
angle in previous studies (Su et al. 2007; Aulanier et al. 2012).
Similar to the pre-flare morphology (Figure 2(i)), the PFLs also
exhibited two different magnetic connectivities. One set of PFLs
seemed to connect ribbons R2 and R3 (red symbols and red dash-dotted
line in panels (a)-(b)), and another overlying set of PFLs
connecting R1 and R3 (blue symbols and blue dash-dotted line in
panels (a)-(c)). In Figures 4(d)-(e), we measured the mutual
orientation between the two sets of PFLs (blue dotted lines
connecting ribbons R1 and R3; red dotted lines connecting R2 and
R3). The angles (40, 60 and 65$^{\circ}$) are obtained by measuring
the angles between the tangents of red and blue loops. Starting from
23:30 UT, about 35 PFLs have been identified and their $\theta$
values were estimated. Figure 5 shows the results of $\theta$ for
these PFLs, separated into two groups according to their
connectivity. The set of PFLs connecting R1 and R3 has a higher
non-potentiality, e.g., strong shear (or deviating from potential
fields). Their $\theta$ values range from 15$^{\circ}$ to
67$^{\circ}$ (blue symbols), and the average $\theta$ is about
48$^{\circ}$ (see Table 1). Another set of PFLs connecting R2 and R3
has larger $\theta$ values of 63$^{\circ}$-90$^{\circ}$, implying
that they are approximately perpendicular to the PIL and weakly
sheared fields.

Figure 6 displays the coronal magnetic field lines and the
photospheric Q-map. The extrapolation results show the existence of
two sets of magnetic systems (MS1 and MS2, cyan and blue lines in
panel (a)) and the underlying three sets of core sheared arcades
(CSAs, pink lines in panel (a)) at the flaring region. MS1 seems to
connect ribbons R1 and R3, and MS2 connects ribbons R2 and R3. The
surface delineated by MS1 lies below the corresponding surface of
MS2, with both of them overlying the CSAs. The three CSAs are
strongly sheared with the average twist number of about 0.5-0.7,
which probably correspond to the three non-eruptive filaments (F1-F3
in Figure 2(a)). We identified four field line strands (FL1-FL4 in
Figure 6(b)) by comparing with the observed high-temperature loop
bundles (L1-L4 in Figure 2(i)), and plotted them over the
photospheric Q-map (Figure 6(b)). The four strands FL1-FL4 are
anchored in the locations with high Q value of about
10$^{4}$-10$^{5}$, indicating the positions of QSLs with strong
connectivity gradients. FL1-FL2 delineate a QSL structure labeled
Q1, and FL3-FL4 outline another QSL structure Q2 (Figure 6(b)). The
overlay between the flare ribbons (underlying 1600 {\AA} image) and
the photospheric Q-map is shown in panel (c). We can see that the Q
distribution nearby ribbon R1 is very complex and is poorly matched
with R1. However, the other two ribbons R2 and R3 have a good
correspondence with high Q regions, with R2 residing in the north
end of Q2 and R3 in the common west end of Q1 and Q2. The close
relations between observed flare ribbons and calculated high Q
regions imply that the dominant reconnection process during the
flare occurs along the two QSLs. It is noted that not all high-Q
regions have corresponding flaring activity. The reason is that in
some high-Q regions, no current is accumulated and thus no flaring
activity is observed (Savcheva et al. 2015). In order to estimate
the magnetic field gradient in the environment of CSAs, we computed
the distribution of the decay index n above the PIL prior to the
flare onset (Figure 6(d)). The black line marks the position where n
reaches the critical value of 1.5 for the onset of torus instability
(Kliem \& T{\"o}r{\"o}k 2006). The critical height is above 45 Mm at
all portions above the PIL.

\subsection{``Type I": the X2.0-class Flare on 2014 October 26}

Another selected event of ``type I" confined flares is the
X2.0-class flare occurring in the famous AR 12192 on 2014 October
26. AR 12192 was the biggest sunspot region in the solar cycle 24
and produced 6 X-class flares, 22 M-class flares, and 53 C-class
flares during its disk passage. The most peculiar aspect of this AR
was that all the X-class flares were confined and none of them were
associated with CMEs. The famous AR drew considerable attention and
has been extensively studied (Thalmann et al. 2015; Sun et al. 2015;
Chen et al. 2015; Liu et al. 2016a; Sarkar \& Srivastava 2018). They
suggested that the weaker non-potentiality and stronger strapping
magnetic field resulted in the confinement of the flares. Zhang et
al. (2017) analyzed the evolution of four confined X-class flares on
2014 October 22-26 and concluded that the complex magnetic
structures are responsible for the confined character of solar
flares.

A total of 6 flares in AR 12192 satisfied our selection criteria,
including 4 X-class flares and 2 M-class flares. The X2.0-class
flare on October 26 was analyzed in detail in this study, which has
never been investigated thoroughly in previous studies. The
X2.0-class flare has the simultaneous high-resolution observations
from the \emph{Interface Region Imaging Spectrograph} (\emph{IRIS};
De Pontieu et al. 2014) showing clear dynamic evolution of flare
ribbons. The onset time of the X2.0-class flare was 10:35 UT and the
peak time was 10:56 UT as shown from the GOES SXR 1$-$8 {\AA} flux
(Figure 8(e)). The AR was located at the S$10^{\circ}$-$20^{\circ}$
latitude and W$30^{\circ}$-$45^{\circ}$ longitude during the flare.
The 304 {\AA} observations showed that a ``reverse S-shaped"
filament was present along the PIL of the AR, with a length of about
45 Mm (green arrows in Figure 7(a)). After the flare onset, the
western part of the filament was activated and associated with EUV
brightenings (Figure 7(b)). However, the entire filament did not
show any rise phase except for the moderate brightenings and
remained stabilized through the flare. Three flare ribbons appeared
in the central region of the AR, including one negative-polarity
ribbon (R1) in the trailing sunspot of the AR and two
positive-polarity ribbons (R2 and R3) anchoring in the periphery of
the leading sunspot (Figures 7(c)-(d)). The location and morphology
of the flare ribbons in this event are similar to the other five
flares in the same AR, implying that they are homologous flares with
an analogous triggering mechanism.

High-temperature flare loops exhibited a complicated structure as
seen from the 131 and 94 {\AA} observations. Four loop bundles
(L1-L4 in panels (e)-(g)) overlying the non-eruptive filament were
identified, which were probably involved in the magnetic
reconnections of the flare. At 10:28 UT prior to the flare, L1 and
L2 showed faint brightenings (panel (e)), indicative of the onset of
weak magnetic reconnections. Then at 10:38 UT, a shorter and
brighter loop bundle L3 (panel (f)) appeared underlying L1 and L2.
Simultaneously, three flare ribbons R1-R3 were formed at the
footpoints of the heated flare loops (panel (f)). Associated with
the development of the flare, flare loops connecting ribbons R1 and
R2 exhibited an apparent slipping motion towards the north along
ribbon R1 (red arrow in panel (g)), implying the occurrence of
slipping magnetic reconnections. At the peak time of the flare, the
southernmost loop L4 connecting ribbons R1 and R3 was seen in the 94
{\AA} channel, and L3-L4 jointly comprised a ``triangle-shaped flag
surface" (Figures 7(c) and (g)). At last, these high-temperature
flare loops gradually cooled down and formed PFLs overlying the
non-eruptive filament in 171 {\AA} (panel (h)). We suggest that
L1-L2 outline a set of sheared magnetic system and L3-L4 delineate
another set of magnetic system. The two systems are interacting and
reconnecting with each other, which generates the confined flare.

This flare was also observed by the \emph{IRIS} slit jaw imagers
(SJIs) in 1330 and 2796 {\AA} channels with a spatial pixel size of
0$\arcsec$.33, a field of view (FOV) of
120$\arcsec$$\times$119$\arcsec$ and a cadence of about 18 seconds.
Figure 8 shows the observational results from the \emph{IRIS},
displaying the detailed dynamic evolution of flare ribbons R1 and
R3. As seen from the stack plot (Figure 8(e)) along slice ``S3"
(dash-dotted line in Figure 8(a)) in 1330 {\AA}, ribbons R1 and R3
spread fast perpendicular to the PIL at respective speeds of about
11 km s$^{-1}$ and 21 km s$^{-1}$ along the same direction. As seen
from Table 1, four of six flares in the same AR exhibited the
perpendicular motion with respect to the PIL at speeds of 10-21 km
s$^{-1}$. The ribbon motion is probably controlled by the magnetic
environment and the ribbons in some flares can not really move due
to strong fields.

In addition to the motions perpendicular to the PIL, bidirectional
slipping motion of ribbon R1 along the PIL was also detected (orange
and blue arrows in panel (b)). To analyze the slipping motions, we
place an hook-shaped cut ``S4" (dashed curve in panel (b)) along
ribbon R1 and obtained the stack plots in \emph{IRIS} 1330 and 2796
{\AA} (Figures 8(f)-(g)). The slipping motions were in both
directions with speeds of 10-20 km s$^{-1}$. These velocities are
generally lower than those reported from other flares (Dud{\'{\i}}k
et al. 2014, 2016; Li \& Zhang 2015; Li et al. 2018b). The overlay
of the flare ribbon time evolution over the magnetogram is displayed
in Figure 8(c). The color indicates the time of the ribbon
brightness observed in \emph{IRIS} 1330 {\AA} SJIs. It clearly shows
the bidirectional elongations of ribbon R1 and the unidirectional
perpendicular expansions of ribbons R1 and R3. Figures 8(d1)-(d6)
display the zoomed 2796 {\AA} images of the south part of ribbon R1.
As seen from these zoomed images, ribbon R1 was composed of numerous
bright knots, which exhibited apparent slipping motions along R1.
Three individual bright knots within R1 were tracked (labeled as
``1"$-$``3") at 10:42-10:47 UT. Bright knot ``1" slipped towards the
east along the straight part of R1, with a displacement of about 2.4
Mm in 1.5 min and an average speed of 27 km s$^{-1}$. Bright knot
``2" slipped toward the west at a faster speed of about 36 km
s$^{-1}$. At 10:45:34 UT, another bright knot ``3" was traced to
slip in the same direction as knot ``1". The elongation motion of
flare ribbons parallel to the PIL can also be observed in other five
flares of the same AR (see Table 1). The elongation velocity is in
the range of 11-45 km s$^{-1}$, which is comparable to the previous
case and statistical studies (Krucker et al. 2003; Lee \& Gary 2008;
Qiu et al. 2017).

Starting from about 11:00 UT, the reconnected flare loops gradually
cooled down and formed PFLs in 171 {\AA} (Figures 9(a)-(c)). A set
of PFLs (PFL1 in Figure 9(a)) connecting the south part of ribbon R1
and R3 were initially detected. We estimated the inclination angles
$\theta$ of PFL1 with respect to the PIL and displayed the
measurement results in Figure 10. The PIL in this event is
determined according to the position of the non-eruptive filament
(Su et al. 2006). PFL1 shows an increasing shear, with the $\theta$
values ranging from 70$^{\circ}$ to 30$^{\circ}$ (red symbols in
Figure 10). As seen from the 304 {\AA} images, PFL1 appeared as
bright and dark alternatively arcades (Figure 9(d)), indicative of
the presence of hot and cold materials along the PFLs. PFL1 were
overlying the non-eruptive filament (green arrows in Figure 9(d)),
implying that the filament did not play any part in the reconnection
process. From about 11:28 UT, another set of PFLs (PFL2 in Figures
9(b) and (e)) appeared connecting the north part of ribbon R1 and
R3. The inclination angles $\theta$ of PFL2 were in the range of
30$^{\circ}$-50$^{\circ}$ (purple symbols in Figure 10), suggesting
that PFL2 were strongly sheared loops with respect to the PIL. About
1 hour after the flare peak (11:50 UT), another set of large-scale
PFLs were observed connecting ribbons R1 and R2 (PFL3 in Figures
9(c) and (f)). They were nearly parallel to the PIL and had a higher
non-potentiality, with the $\theta$ of 10$^{\circ}$-25$^{\circ}$
(blue marks in Figure 10). The three sets of PFLs were successively
formed and ultimately delineated two groups of magnetic
connectivities (Figure 9(f)), similar to the pre-reconnecting
high-temperature flare loops (Figure 7). Moreover, the PFLs of the
other five flares in the same AR also exhibited a strong shear, with
$\theta$ values of 20$^{\circ}$-45$^{\circ}$ (see Table 1).

Figure 11(a) shows the 3D structure of the magnetic field lines of
the NLFFF based on the photospheric vector magnetogram at 09:46 UT.
We find that a set of CSAs along the PIL and two sets of sheared
magnetic systems (MS1 and MS2) overlying the flux rope are present
in the central region of the AR. CSAs consist of weakly twisted
field lines with the average twist number of 0.6. Compared to Figure
7, CSAs bear a good resemblance to the observed non-eruptive
filament. The east ends of MS1 and MS2 both anchored in ribbon R1
and their respective west ends in two positive-polarity ribbons R2
and R3. The extrapolated magnetic topology of the AR core is
approximately consistent with the results of Inoue et al. (2016) and
Jiang et al. (2016), who analyzed the X3.1-class flare on October 24
and suggested that the AR was composed of multiple strongly-sheared
flux tubes. Based on the observed loop bundles L1-L4 in Figure 7, we
select four strands of field lines (FL1-FL4 in Figure 11(b)) and
find that their footpoints are located at the regions with high Q
value. Thus it is deduced that two QSLs are connected with the
flare, which are respectively outlined by FL1-FL2 and FL3-FL4.

The photospheric intersections of the two QSLs (Q1 and Q2) and the
brightening ribbons in 1600 {\AA} image (panel (c)) show an
approximate correspondence. Ribbon R1 has a similar morphology with
the common eastern end of Q1 and Q2, although there is a little
displacement between them. The displacement is probably caused by
the evolution of QSL structures during the development of the flare.
The western two ribbons R2 and R3 are approximately matched with the
western ends of Q1 and Q2. Due to the evident evolution of flare
ribbons perpendicular to the PIL (Figure 8), the correspondences
between pre-flare QSLs and flare ribbons are not as good as the
first event on 2011 March 09. We display the distribution of the
decay index n above the PIL prior to the flare onset in panel (d).
It shows that the decay index n does not reach the critical value of
1.5 until 90 Mm, implying a strong confinement overlying the
filament.

\subsection{``Type II": the M5.3-class Flare on 2012 July 04}

One selected event of ``type II" flares is the M5.3-class flare
occurring in AR 11515 on 2012 July 04. The GOES soft X-ray 1-8 {\AA}
flux showed that the flare initiated at 09:47 UT and reached its
peak at 09:55 UT (Figures 15(g)-(h)). The \emph{SDO}/HMI LOS
magnetograms showed that the photospheric magnetic field at the
flaring region has a tripolar structure in which the
negative-polarity patch N1 emerged between positive-polarity sunspot
P1 and positive-polarity patch P2 (Figure 12). The emergence of N1
started almost from the beginning of July 03, and simultaneously N1
and P2 exhibited the shearing motion in opposite directions (red and
blue arrows in Figures 12(a)-(c)). More significantly, the
positive-polarity sunspot P1 showed a converging motion towards the
PIL between P1 and N1 (orange arrows in panels (a)-(c)). Along the
converging direction (``S5" in panel (a)), we obtained a stack plot
based on HMI LOS magnetograms and found that the converging speed of
P1 towards the south was about 0.1 km s$^{-1}$ (panel (e)). As seen
from the 304 {\AA} image, a filament was present along the PIL
between P1 and N1 (panel (d)), which erupted later on and generated
the M5.3-class flare. The filament had a ``reverse S" shape and
seemed to be composed of multiple twisted fine structures.

The 304 {\AA} observations showed that the southwest part of the
filament started to rise up slowly from about 09:45 UT and the
remanent part was still stable (Figure 13(a)). The ascent of the
filament was associated with the EUV brightenings near its two ends
(Figure 13(b)). To study the kinematics of the filament in detail,
we take a slice along the eruption direction of the filament (``S6"
in Figure 13(i)). Figures 14(d)-(e) show the stack plots of the
slice in 131 and 335 {\AA} passbands. The evolution of the filament
exhibited a slow rise phase and a rapid-acceleration phase. The
initial speed of the filament was $\sim$10 km s$^{-1}$ during the
slow rising process at about 09:45-09:49 UT (Figure 14(d)). The
flare was initiated at 09:47 UT, about 2 min later than the slow
rise of the filament. Almost from 09:50 UT, the erupting velocity
increased rapidly and reached $\sim$100 km s$^{-1}$. The 335 {\AA}
stack plot shows a larger velocity of $\sim$150 km s$^{-1}$ in the
impulsive acceleration phase (Figure 14(e)). Later on, the velocity
of the filament started to decrease. Finally the filament material
drained back along its west leg to the solar surface (Figures
13(c)-(d) and 14(d)-(e)), and the eruption became failed (Ji et al.
2003). As seen from the 171 and 94 {\AA} images, large-scale EUV
loops were present over the flaring region (L1 and L2 in Figures
13(e)-(h)). Associated with the eruption of the filament, these
large-scale EUV loops were disturbed and pushed outward. Taking L2
for example, its projected height increased by about 24 Mm in 6 min.
The flare consisted of two main ribbons (R1 and R2 in Figure 13(k))
and two weakly-brightened secondary ribbons (SR1 and SR2).

Two main ribbons R1 and R2 do not exhibit discernable separation
motion perpendicular to the PIL, probably due to the block effect of
the strong magnetic field in the sunspot. We also do not see ribbon
elongations along the PIL, implying the almost simultaneous magnetic
reconnections along the PIL. The pre-eruption filament, four flare
ribbons and large-scale loop bundles are plotted over the LOS
magnetogram (Figure 13(l)) to analyze their magnetic connectivity.
It showed that the filament was located between two main ribbons R1
and R2, implying that the main reconnection process occurs
underlying the eruptive filament. R1 and R2 respectively anchored in
the converging positive-polarity sunspot P1 and the shearing
negative-polarity patch N1 (Figures 12 and 13(l)). Large-scale loop
bundle L1 connected two secondary ribbons SR1 and SR2, indicative of
their conjugated property. SR1 was located at the southernmost
shearing positive-polarity patch P2 and SR2 at the remote
negative-polarity sunspot (Figures 12 and 13(l)). Loop bundle L2 was
seen to connect the positive and negative sunspots of the AR, which
probably constrained the eruption of the filament.

Starting from about 10:00 UT, PFLs appeared underlying the eruptive
filament due to the cooling of reconnected loops (Figures 13(g)-(h)
and 14(b)-(c)). These PFLs seemed to be quasi-parallel with each
other, connecting two main ribbons R1 and R2. As seen from Figure
14(a), the PIL between ribbons R1 and R2 is strongly curved
encircling the positive-polarity sunspot. To evaluate the
non-potentiality of the PFLs, we measured the inclination angles
$\theta$ of PFLs with respect to the PIL (Figures 14(b)-(c)). The
$\theta$ value is about 80$^{\circ}$-86$^{\circ}$, implying that the
PFLs are approximately perpendicular to the PIL and nearly potential
fields.

In the decay phase of the flare, the central flare loops gradually
faded away, however, another set of longer loops connecting the
central flaring region with the remote ribbon started to brighten up
(Figures 15(a)-(b)). These brightening loops initially appeared in
the high-temperature passbands such as 131 and 94 {\AA} (Figures
13(i)-(j) and 15(a)-(b)), then became visible sequentially in cooler
AIA passbands such as 335 (about 2.5 MK) and 171 {\AA} (about 0.6
MK; Figures 15(c)-(f)). These loops are morphologically similar in
different passbands, implying that they are the same structures. We
cut a slice of the AIA 335 {\AA} images (``S7" in Figure 15(d)) and
plotted its time evolution in Figure 15(g). It showed that these
longer brightening loops were formed at about 10:45 UT in 335 {\AA},
with a time delay of 50 min after the flare peak. Thus the long
brightening loops are identified as late-phase loops (Woods et al.
2011; Liu et al. 2013). About 60 min later, the late-phase loops
(LPLs) in 335 {\AA} started to cool down at 11:45 UT. The EUV
emissions summed over the cutout of the AR (white rectangle in panel
(b)) in different passbands are shown in Figure 15(h). All the EUV
emission variations in different passbands exhibit a main phase and
a late phase. The emission variations in all temperatures reach
their peaks at almost the same time in the main phase, however there
are larger time lags between the peaks of the late phase in
different temperatures. In 94 {\AA}, the peak flux of the late phase
is almost the same as the main phase and the time lag between the
two peaks is about 40 min. The time differences between the late
phase peaks and the main phase peaks in 335 and 171 {\AA} are 80 and
85 min, respectively.

The topology of the 3D magnetic field reveals the existence of a
flux rope (FR) and the overlying constraining fields (Figure 16(a)).
FR is moderately twisted with an average twist number of 2.0, and
bears a good resemblance to the observed filament (Figures 12 and
13). The overlying cyan fields connect the flaring region with
remote brightenings, corresponding to the identified large-scale
loop bundles L1 and L2 in Figure 13. As seen from the distribution
of the Q factor (Figure 16(b)) in a vertical plane across the
pre-eruptive FR axis (yellow bar in Figure 16(a)), a QSL structure
(Q1) of upside-down teardrop shape at the boundary of the FR and a
larger dome-shaped QSL (Q2) encircling Q1 are present. In panel (c),
we show the matches between the QSLs and the flare ribbons. It is
seen that the flare ribbons are well matched with the two QSLs (Q1
and Q2). At the locations of two strongly curved main ribbons R1 and
R2, the intersections of Q1 with the lower boundary are present and
display the similar shapes with ribbons R1 and R2. This implies that
magnetic reconnection probably occurs at the FR-related Q1
underlying the FR and results in the formation of main ribbons R1
and R2 (Janvier et al. 2014; Jiang et al. 2018; Liu et al. 2018a),
which is consistent with the CSHKP flare model. The majority of
secondary ribbon SR1 is well matched by the southern footpoints of
the dome-shaped Q2. We note that the west hook of SR1 is poorly
matched, probably due to the higher complexity of magnetic fields at
this location. The remote secondary ribbon SR2 shows a good
correspondence with the remote footpoints of Q2 (eastern ends of
cyan lines in panel (a)), indicative of the occurrence of secondary
magnetic reconnection along the large-scale Q2.

In Figure 16(d), we display the distribution of the decay index n
above the PIL prior to the flare onset. It is seen that the decay
index n shows an unusual distribution. Above the eastern half of the
PIL (30-60 Mm along the PIL), n reaches 1.5 at varying heights,
which are all above 90 Mm, suggesting the presence of strong
confinement at this region. Above the western half of the PIL (0-30
Mm along the PIL), n reaches 1.5 at a height around 15 Mm, keeping
larger than 1.5 until 40 Mm, then drops below 1.5 in a large range
of height (from 40 Mm to a height larger than 150 Mm), though n is
increasing in this height range. This kind of distribution is called
a ``saddle-like" profile, which has been studied (Guo et al. 2010;
Wang et al. 2017; Liu et al. 2018b), and is usually associated with
failed eruption. The ``saddle-like" profile exhibits a local
torus-stable (n$<$1.5) region enclosed by two torus-unstable
domains. Eruptions occurring in a region having ``saddle-like" decay
index distribution may be slowed down in the highly located
torus-stable region if the initial disturbance is not large enough,
thus fails to erupt out into the interplanetary space. In our case,
the eruption of the filament occurred at the western part of the
filament (see Figure 13), above which n had a ``saddle-like"
distribution, having an apparent rising velocity around 150 km
s$^{-1}$ (Figure 14). The eruption may happen when the flux rope
supporting the filament enters the local torus-unstable region
between 15-40 Mm. It keeps rising up and then enters the
torus-stable region. This region provides strong confinement again.
With a small initial velocity, the flux rope has no enough
disturbance to erupt out and is slowed down by the slow decaying,
strong strapping fields in the torus-stable region. The particular
distribution of n, along with the clear signature of a failed
eruption of a filament, suggests that the strong confinement above
the filament plays the major role in confining the eruption in this
case.

\section{Discussion}

The two types of flares display different flaring structures and
dynamic evolution. We estimated the inclination angles $\theta$ of
PFLs with respect to the PIL for 15 flares (PFLs in 3
circular-ribbon flares are too compact and thus are not measured in
Table 1) and displayed the histogram for the two types of flares in
Figure 17. It shows that the $\theta$ for ``type I" is in the range
from $10^{\circ}$ to $50^{\circ}$. The $\theta$ for ``type II" is
evidently higher than ``type I", in the range of $60^{\circ}$-
$90^{\circ}$. The difference of the $\theta$ value for the two types
of flares is probably caused by the different magnetic environments
where reconnection occurs. For ``type I" flares, slipping
reconnection occurs between two sets of magnetic systems and results
in the interchange of their magnetic connectivities. The mutual
orientation between two reconnecting systems is less favorable for
reconnection, and thus the reconnection is limited and the new
reconnected magnetic field is still strongly sheared (Galsgaard et
al. 2007; Zuccarello et al. 2017). For ``type II" flares, magnetic
reconnection occurs in anti-parallel magnetic fields underlying the
erupting flux rope or at the 3D null-point in the fan-spine
topology, which is the effective reconnection process leading to the
flux rope erupiton (Archontis \& T{\"o}r{\"o}k 2008; Leake et al.
2013, 2014). In this situation, a large majority of free magnetic
energy is released and the newly formed PFLs generally relax fully
to a quasi-potential state.

In order to estimate the magnetic field gradient in the environment
of the filaments, we computed the distribution of the decay index n
above the PIL. For the first flare on 2011 March 09, the decay index
distribution has a modest critical height around 45 Mm (Figure 6).
While for the event occurring on 2014 October 26, the distribution
has a critical height above 90 Mm (Figure 11). The former is a
modest height with which both confined and eruptive flares may occur
(e.g., Liu et al. 2016a). The latter seems to indicate a strong
confinement overlying the filament. However, the magnetic
reconnections during the two ``type I" flares occur in these
strapping fields overlying the filaments (MS1 and MS2 in Figures
6(a) and 11(a)), so it is difficult to distinguish the overlying
fields as reconnecting fields or confining fields. In both events,
the filaments are stable and do not show any eruption signatures.
Thus the confinement of the overlying strapping fields, no matter it
is large or small, would not play a significant role in the
eruptiveness of the flares. We suggest that the analysis of the
decay index in complex ``type I" flares does not provide a useful
clue for the class of eruption (Zuccarello et al. 2017).

For the ``type II" flare on 2012 July 04, the decay index
distribution has a ``saddle-like" profile at the western part above
the filament (Figure 16), having a local torus-unstable region
within 15-40 Mm, which is enclosed by two torus-stable regions. The
filament experienced a failed eruption at the western part of the
PIL with a relatively slow initial velocity around 150 km s$^{-1}$
(Figures 13-14). With this small initial disturbance, the flux rope
supporting the filament may be slowed down in the higher located
torus-stable region, thus fails to erupt out. Unlike ``type I"
confined flares, this ``type II" event has clear failed eruption
signatures, and its post-flare loops have weakly sheared
configuration, suggesting that most helicity of the erupting system
is released during the eruption. Thus, a flux rope must have erupted
at first, though finally being stopped in the higher corona. The
strong confinement of the strapping magnetic fields should have
played a significant role in confining the eruption associated with
this kind of standard confined flares.

All the ``type I" flares exhibit the elongation motion parallel to
the PIL at speeds of a few tens of km s$^{-1}$ (Table 1). The
elongation motion of flare ribbons is not present in a 2D framework,
however, it can be explained by the 3D reconnection along the QSLs
(Priest \& D{\'e}moulin 1995; Masson et al. 2012). High-resolution
observations of the \emph{IRIS} displayed the bi-directional
slipping motion of ribbon substructures (Figure 8). We suggest that
the slipping substructures along two directions respectively
correspond to the footpoints of two magnetic systems (MS1 and MS2 in
Figure 11). The continuous slipping magnetic reconnection between
the two magnetic systems results in the exchange of their
connectivities and the tangential movement of reconnecting field
lines one to each other. Flare loops in this kind of flares also
displayed the apparent slipping motions along the ribbons. In the
first event, the acceleration process of the slipping motion during
the flare is first presented (Figure 3). The slippage of flare loops
exhibits two kinematic phases: a slow slipping phase and a fast
slipping phase. The simulation results of Janvier et al. (2013)
showed that the slipping speed of field lines would increase as the
expansion of the torus-unstable flux rope resulted in the evolution
of QSLs toward separatrices. The acceleration of the slippage in our
observations implies that the QSLs are displaced devoid of any flux
rope eruption. All these observations of flare loops or ribbon
motions during ``type I" confined flares provide the evidence for
slipping magnetic reconnections between different sets of magnetic
systems. The constant flux emergence and shearing motion (Figure 1)
probably help to accumulate the current at the interface between the
two systems (Krall et al. 1982; Yan et al. 2018). Magnetic
reconnections could occur within the current layer in the vicinity
of QSLs, similar to the simulations of Aulanier et al. (2005). The
slipping reconnection causes no significant topological change and
thus the equilibrium of the entire system is not destabilized. The
flares finally develop into confined events due to the lack of any
erupting magnetic structures such as magnetic flux ropes and sheared
magnetic loop bundles. The scenario of ``type I" confined flares is
similar to the ``unorthodox" and ``atypical" flares in previous case
studies (Liu et al. 2014; Dalmasse et al. 2015; Joshi et al. 2019),
which are both inconsistent with the classical flare models.

``Type II" confined flares are accompanied by the flux rope
eruption, that becomes failed due to the presence of strong
strapping fields overlying the flaring region. They can be described
by the classical flare models. The two-ribbon ``type II" flares are
consistent with the CSHKP model (Shibata \& Magara 2011) and its
extension in 3D (Aulanier et al. 2012; Janvier et al. 2014). The
circular-ribbon ``type II" flares have a fan-spine topology and the
null-point reconnections lead to the occurrence of the flares (Lau
\& Finn 1990; Priest \& Titov 1996; Liu et al. 2011). The early
dynamic evolution of the ``type II" flare is similar to eruptive
flares, differently, the material motion of the filament is hindered
and its trajectory is changed while encountering with the ambient
background fields in the gradual phase of the flare. This kind of
confined flares has been extensively studied and the strength of the
overlying field is thought to be an important factor determining
whether a flare is confined (Cheng et al. 2011; Nindos et al. 2012;
Liu et al. 2016a; Amari et al. 2018). In the event on 2012 July 04,
the formations of secondary flare ribbons and late-phase flare loops
(Figures 13 and 15) both suggest that the large-scale constraining
fields overlying the erupting flux rope are partially reconnected
(Sun et al. 2013; Dai \& Ding 2018). However, the remaining
constraining fields that are not reconnecting still hold a high flux
ratio compared to the flux rope, hence inhibit the eruption of the
flux rope.

\section{Conclusion}

In this study, we selected 18 confined flares (GOES class $\geq$M5.0
and $\leq$$45^{\circ}$ from disk center) that occurred between 2011
January and 2017 December, i.e., 7 yr from the activity minimum of
solar cycle 24. According to their different dynamic properties and
magnetic configurations, we first divide the confined flares into
two types. For ``type I" confined flares, the magnetic configuration
is very complex with two or more QSLs overlying the core magnetic
structure, and multiple slipping reconnections along these QSLs
trigger the occurrence of the flare. Eventually, the slipping
magnetic reconnections do not cause any eruption of magnetic
structures and the entire magnetic system still remains stabilized.
``Type II" confined flare has the magnetic configuration consistent
with the classical flare models, but strong strapping fields are
present over the flaring region in the high atmosphere. Based on the
classification criteria, 7 flares of 18 ($\sim$39 \%) belong to
``type I" and 11 ($\sim$61 \%) are ``type II" confined flares.

The complexity of the magnetic fields involved in ``type I" flares
is shown in several aspects. The pre-flare loops and PFLs are both
composed of two or more sets of magnetic systems. PFLs have a
stronger non-potentiality, with the inclination angle $\theta$ in
the range of 10$^{\circ}$-50$^{\circ}$. All the ``type I" flares
exhibit the ribbon elongations parallel to the PIL at speeds of
several tens of km s$^{-1}$. All these observations indicate the 3D
nature of magnetic reconnection in solar flares. We suggest that
magnetic reconnection between different magnetic systems results in
the flare occurrence. However, the reconnection is probably limited
and has a low efficiency for the mutual orientation between two
systems is less favorable for reconnection. Thus the equilibrium of
the entire system is not destabilized and the flare finally develops
into a confined event.

The two types of confined flares show distinct properties in several
aspects. The complex PFLs in ``type I" flares are composed of two or
more sets of magnetic connectivities overlying the non-eruptive
filament and are strongly sheared. However, the PFLs in ``type II"
flares are formed underlying the erupting magnetic structure. They
are approximately parallel with each other and have a weak
non-potentiality. Moreover, all the ``type I" flares exhibit the
ribbon elongations along the PIL at apparent speeds of several tens
of km s$^{-1}$, suggestive of the occurrence of slipping magnetic
reconnections along the QSLs. Only 3 of 10 ``type II" flares display
the ribbon elongations. This implies that unlike ``type I" flares
the QSL reconnection is probably not dominant in ``type II" flares,
and in most ``type II" flares magnetic reconnection is almost
simultaneous along the current sheet, consistent with the CSHKP
model. The filament at the flaring region plays different roles in
these two types of confined events. The filament in ``type I" seems
not to be affected by the flare and is not associated with any
eruption process. In ``type II", the filament becomes unstable due
to the loss of equilibrium or suffering MHD instabilities, and the
rise of the filament causes the subsequent reconnection and the
occurrence of the flare.

Overall, our results suggest that there are two types of confined
flares that are triggered by different physical mechanisms. To our
knowledge, ``type II" confined flares have been extensively studied
in the literature and the main reason for the confinement has been
well understood. However, ``type I" confined flares have been rarely
analyzed due to their complexity. Our study shows that ``type I"
accounts for more than one third of all the large confined flares,
which should not be neglected in further studies.

\acknowledgments {We thank the referee for helpful comments that
improved the paper. \emph{SDO} is a mission of NASA's Living With a
Star Program. \emph{IRIS} is a NASA small explorer mission developed
and operated by LMSAL with mission operations executed at NASA's
Ames Research center and major contributions to downlink
communications funded by the Norwegian Space Center (NSC, Norway)
through an ESA PRODEX contract. This work is supported by the
National Natural Science Foundations of China (11773039, 11533008,
11790304, 11673035, 11673034 and 11790300), Key Programs of the
Chinese Academy of Sciences (QYZDJ-SSW-SLH050), Young Elite
Scientists Sponsors hip Program by CAST (2018QNRC001) and the Youth
Innovation Promotion Association of CAS (2017078). Lijuan Liu was
supported by NSFC (11803096) and the Open Project of CAS Key
Laboratory of Geospace Environment.}

{}
\clearpage

\begin{table}
\setlength{\tabcolsep}{4.0pt}
\caption{Event list}
\label{table2} 

\begin{tabular}{c c c c c c c c c c} 
\hline\hline 
Event & Date & Time\tablenotemark{a} & GOES & AR &
Filament\tablenotemark{b} & Separation\tablenotemark{c} &
Elongation\tablenotemark{d} &
Angle\tablenotemark{e} & Type I/II \\ 
No. & & & class & & & (km s$^{-1}$) & (km s$^{-1}$) & (degree) &  \\
\hline 
1 & 20110309 & 23:23 & X1.5 & 11166 & S & No & 33$\pm$3 & 48$\pm$2 & I \\
2 & 20110730 & 02:09 & M9.3 & 11261 & E & 5$\pm$1 & 11$\pm$2 & 74$\pm$2 & II \\
3 & 20120510 & 04:18 & M5.7 & 11476 & E & No & No & - & II \\
4 & 20120704 & 09:55 & M5.3 & 11515 & E & No & No & 83$\pm$2 & II \\
5 & 20120705 & 11:44 & M6.1 & 11515 & E & No & No & 82$\pm$2 & II \\
6 & 20131101 & 19:53 & M6.3 & 11884 & E & No & 15$\pm$3 & 64$\pm$2 & II \\
7 & 20140107 & 10:13 & M7.2 & 11944 & E & No & No & - & II \\
8 & 20140204 & 04:00 & M5.2 & 11967 & E & No & No & 71$\pm$2 & II \\
9 & 20141022 & 01:59 & M8.7 & 12192 & S & 12$\pm$2 & 45$\pm$5 & 34$\pm$2 & I \\
10 & 20141022 & 14:28 & X1.6 & 12192 & S & No & 16$\pm$4 & 45$\pm$2 & I \\
11 & 20141024 & 21:40 & X3.1 & 12192 & S & 15$\pm$2 & 23$\pm$4 & 25$\pm$2 & I \\
12 & 20141025 & 17:08 & X1.0 & 12192 & S & 10$\pm$1 & 12$\pm$2 & 20$\pm$2 & I \\
13 & 20141026 & 10:53 & X2.0 & 12192 & S & 21$\pm$2 & 20$\pm$3 & 18$\pm$2 & I \\
14 & 20141027 & 00:34 & M7.1 & 12192 & A & No & 11$\pm$3 & 37$\pm$2 & I \\
15 & 20141204 & 18:25 & M6.1 & 12222 & E & 12$\pm$3 & No & 79$\pm$2 & II \\
16 & 20150824 & 07:33 & M5.6 & 12403 & E & - & - & - & II \\
17 & 20150928 & 14:58 & M7.6 & 12422 & E & No & No & 67$\pm$2 & II \\
18 & 20170906 & 09:10 & X2.2 & 12673 & E & 8$\pm$1 & 10$\pm$2 & 73$\pm$2 & II \\
\hline 
\end{tabular}
\begin{list}{}{}
\item[] Notes.
\item[$^{\mathrm{a}}$] Flare peak time.
\item[$^{\mathrm{b}}$] The filament dynamics in the flaring
region: ``S", ``E" and ``A" respectively means stable, eruptive and
activated.
\item[$^{\mathrm{c}}$] The separation motion of flare ribbons
perpendicular to the PIL. The velocity is obtained by the linear
fitting to the stack plot along a slice perpendicular to the PIL. We
assumed an error of two pixels (1$\arcsec$.2) in the location
measurement and the uncertainty in the speed was thus estimated
according to different durations of the separation motions.
\item[$^{\mathrm{d}}$] The elongation motion of flare ribbons along
the PIL. The velocity is obtained by the linear fitting to the stack
plot along a slice along the ribbon. The error estimation is similar
to the separation motion.
\item[$^{\mathrm{e}}$] The inclination angle $\theta$ of PFLs
with respect to the PIL. $\theta$ is the average value by measuring
different PFLs that can be clearly discerned. If there are two or
more sets of PFLs, $\theta$ corresponds to the most strongly sheared
set of loops. The PIL information used in this study is from the LOS
photospheric magnetograms from the \emph{SDO}/HMI.

\end{list}
\end{table}
\clearpage

\begin{figure}
\centering
\includegraphics
[bb=107 204 460 612,clip,angle=0,scale=0.9]{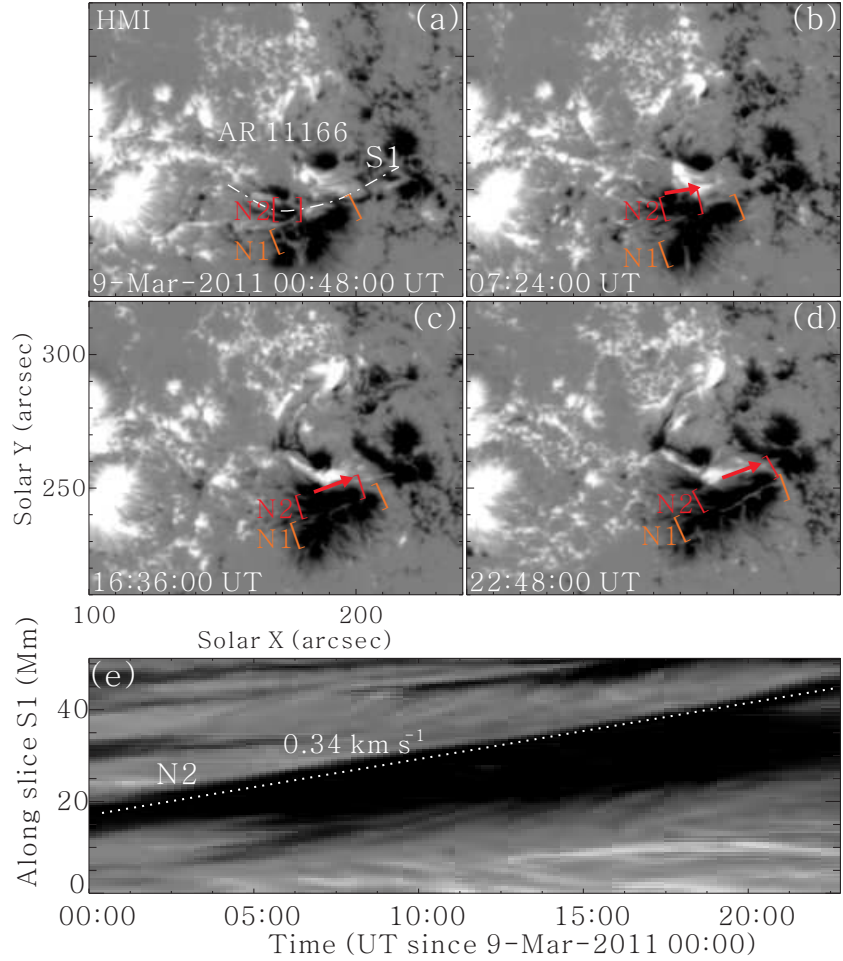}
\caption{Temporal evolution of the HMI LOS magnetic field
(saturating at $\pm$1000 G) prior to the X1.5-class flare on 2011
March 09. Negative patch N1 is the formerly existed magnetic flux
and negative patch N2 is new emerging flux. White dash-dotted curve
``S1" in panel (a) represents the location used to obtain the stack
plot shown in panel (e). Red arrows in panels (b)-(d) indicate the
shearing motion of the emerging flux N2. White dotted line in panel
(e) denotes the shearing motion of N2 along slice ``S1". The FOV of
panels (a)-(d) is the same as Figure 2. \label{fig1}}
\end{figure}
\clearpage

\begin{figure}
\centering
\includegraphics
[bb=18 137 547 680,clip,angle=0,scale=0.8]{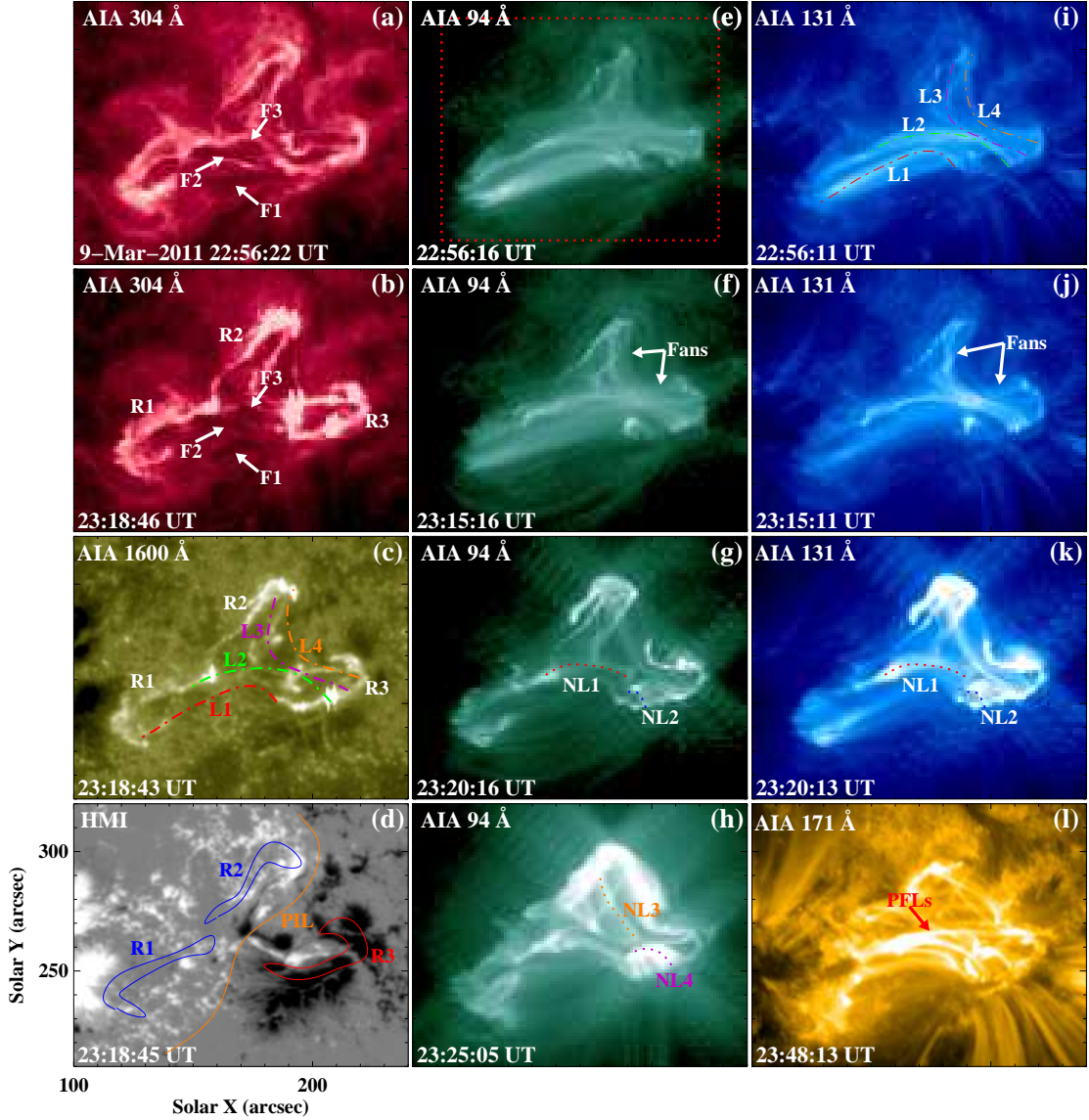}
\caption{Appearance of the X1.5-class flare on 2011 March 09 in
different (E)UV wavelengths and the corresponding LOS magnetogram
from the \emph{SDO}/AIA and \emph{SDO}/HMI. F1-F3 in panels (a)-(b)
are three non-eruptive filaments at the flaring region. R1-R3 in
panels (b)-(d) denote three flare ribbons and they are outlined by
blue and red contours in panel (d). The orange curve in panel (d)
indicates the average orientation of the polarity inversion line
(PIL) of the AR obtained from the HMI LOS magnetogram. The red
dotted rectangle in panel (e) denotes the FOV of Figures 3(a)-(d).
L1-L4 in panels (c) and (i) are four brightened loop bundles
identified in 131 {\AA} prior to the flare. NL1-NL4 in panels
(g)-(h) and (k) denote newly formed loop bundles during the flare.
PFLs in panel (l) represents post-flare loops in the low-temperature
channel of 171 {\AA} after the flare. The animation of this figure
includes AIA 304, 171, 94 and 131 {\AA} images from 22:30 UT to
24:00 UT. The video duration is 37 s. \label{fig2}}
\end{figure}
\clearpage

\begin{figure}
\centering
\includegraphics
[bb=105 173 491 618,clip,angle=0,scale=0.9]{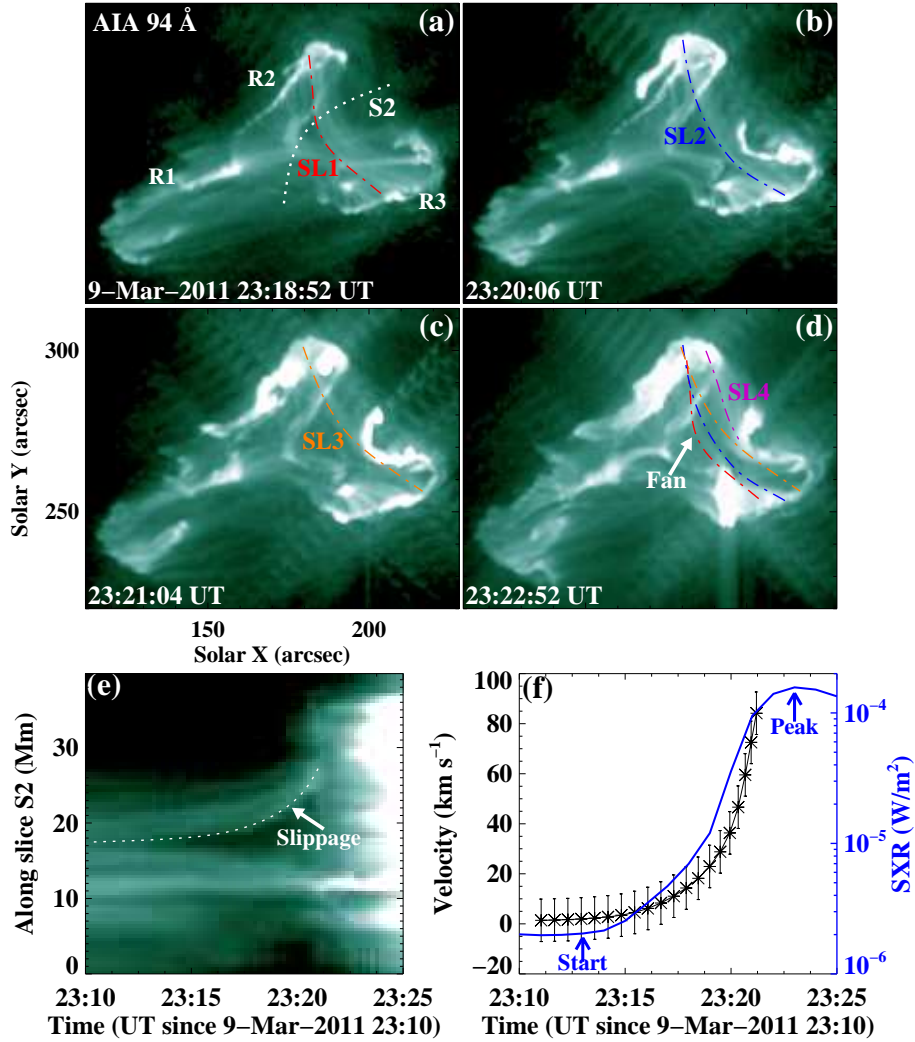}
\caption{Apparent slipping motion of flare loops in the impulsive
phase of the flare. Panels (a)-(d): time series of 94 {\AA} images
showing the slipping flare loops connecting ribbons R2 and R3.
Dash-dotted curves (SL1-SL4) traced the easternmost part of the
slipping loop bundles. These successively visible loop structures
(SL1-SL4) delineated a ``fan-shaped surface" shown in panel (d).
White dotted curve ``S2" in panel (a) denotes the location used to
obtain the stack plot shown in panel (e). Panels (e)-(f): stack plot
along slice ``S2" and the velocity-time profile (black curve)
showing the acceleration process of the slipping motion. The
uncertainty of the speed was estimated to be about 8.5 km s$^{-1}$
due to the uncertainty in the height measurement of 2 pixels. The
blue curve in panel (f) is the GOES SXR 1$-$8 {\AA} flux variation.
\label{fig3}}
\end{figure}
\clearpage

\begin{figure}
\centering
\includegraphics
[bb=20 262 547 553,clip,angle=0,scale=0.8]{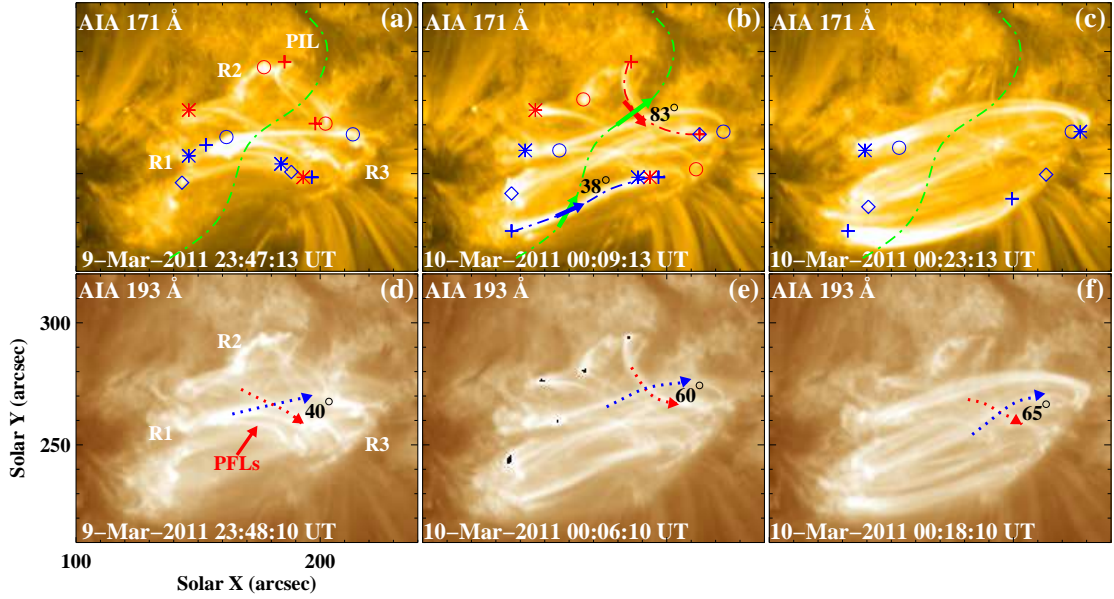} \caption{Time
series of 171 and 193 {\AA} images showing the morphology of PFLs.
Green dash-dotted line is the duplication of the PIL (the orange
curve in Figure 2(d)). 38$^{\circ}$ in panel (b) is the inclination
angle between the blue arrow (the tangent of the blue dash-dotted
line) and the green arrow (the tangent of the PIL). 83$^{\circ}$ is
the inclination angle between the red dash-dotted curve and the PIL.
Blue marks in panels (a)-(c) indicate the footpoints of the measured
PFLs connecting ribbons R1 and R3. Red marks denote the footpoints
of another set of PFLs connecting ribbons R2 and R3. Red and blue
dotted lines in panels (d)-(f) are used to estimate the mutual
orientation between the two sets of PFLs. \label{fig4}}
\end{figure}
\clearpage

\begin{figure}
\centering
\includegraphics
[bb=105 264 477 550,clip,angle=0,scale=0.9]{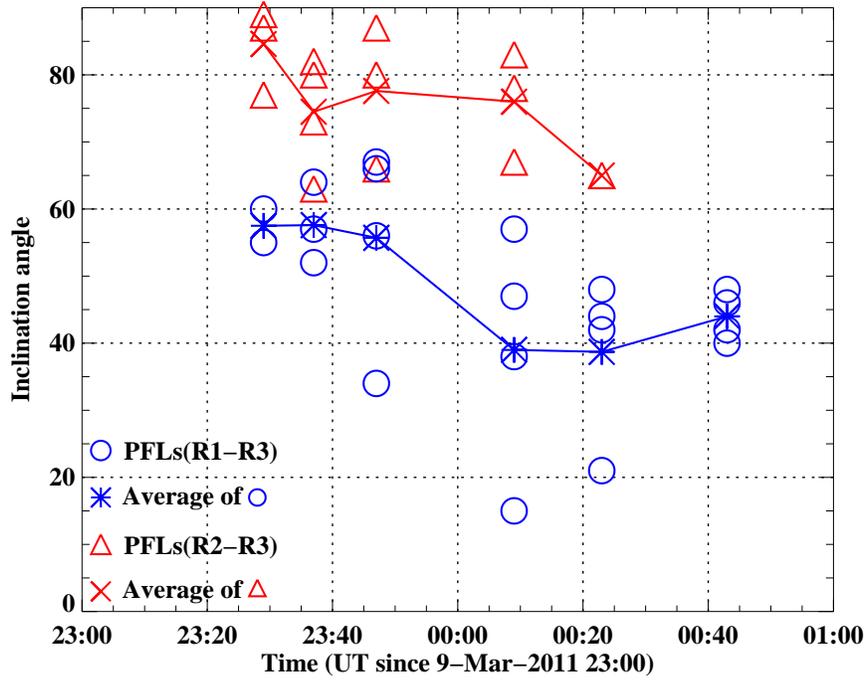}
\caption{Inclination angles of the observed PFLs with respect to the
PIL during the development of the flare. The blue circles denote the
inclination angles of the PFLs connecting ribbons R1 and R3, and the
blue asterisks are the average angles at the same time. The red
symbols are for the PFLs connecting ribbons R2 and R3. \label{fig3}}
\end{figure}
\clearpage

\begin{figure}
\centering
\includegraphics
[bb=130 457 479 760,clip,angle=0,scale=1.0]{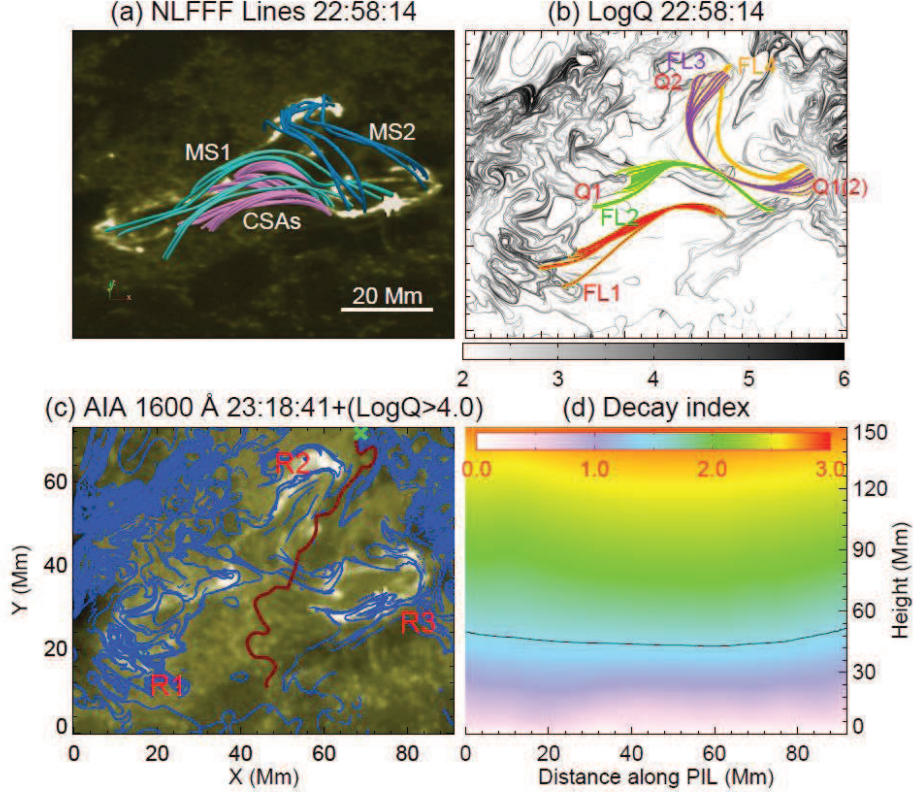} \caption{3D
magnetic field structure, QSL and decay index distribution. Panel
(a): side view of the extrapolated field lines at 22:58 UT showing
two magnetic systems (MS1 and MS2, cyan and blue lines) involved in
the flare and the underlying core sheared arcades (CSAs, pink
lines). The background is the 1600 {\AA} image showing the locations
of flare ribbons. Panel (b): selected four strands of magnetic field
lines (FL1-FL4) by comparing with the observed flare loop bundles
(L1-L4 in Figure 2(i)) plotted over the distribution of the
squashing factor Q on the bottom boundary. FL1-FL2 delineate a QSL
structure Q1 and FL3-FL4 outline another QSL structure Q2. Panel
(c): overlay between the flare ribbons (underlying 1600 {\AA} image)
and the QSL map (blue curves). All regions with
Q$_{thresh}$$>$10$^{4}$ are shown in this panel. The PIL is
extracted from the bottom boundary of the extrapolated NLFFFs and
marked by the red curve. The ``$\times$" symbol shows the starting
point of the PIL. Panels (b)-(c) have the same FOV. Panel (d):
distribution of the decay index above the PIL (red curve in panel
(c)) prior to the flare onset. The black line marks the positions
where decay index n reaches the critical value of 1.5. \label{fig4}}
\end{figure}
\clearpage

\begin{figure}
\centering
\includegraphics
[bb=49 71 508 746,clip,angle=0,scale=0.75]{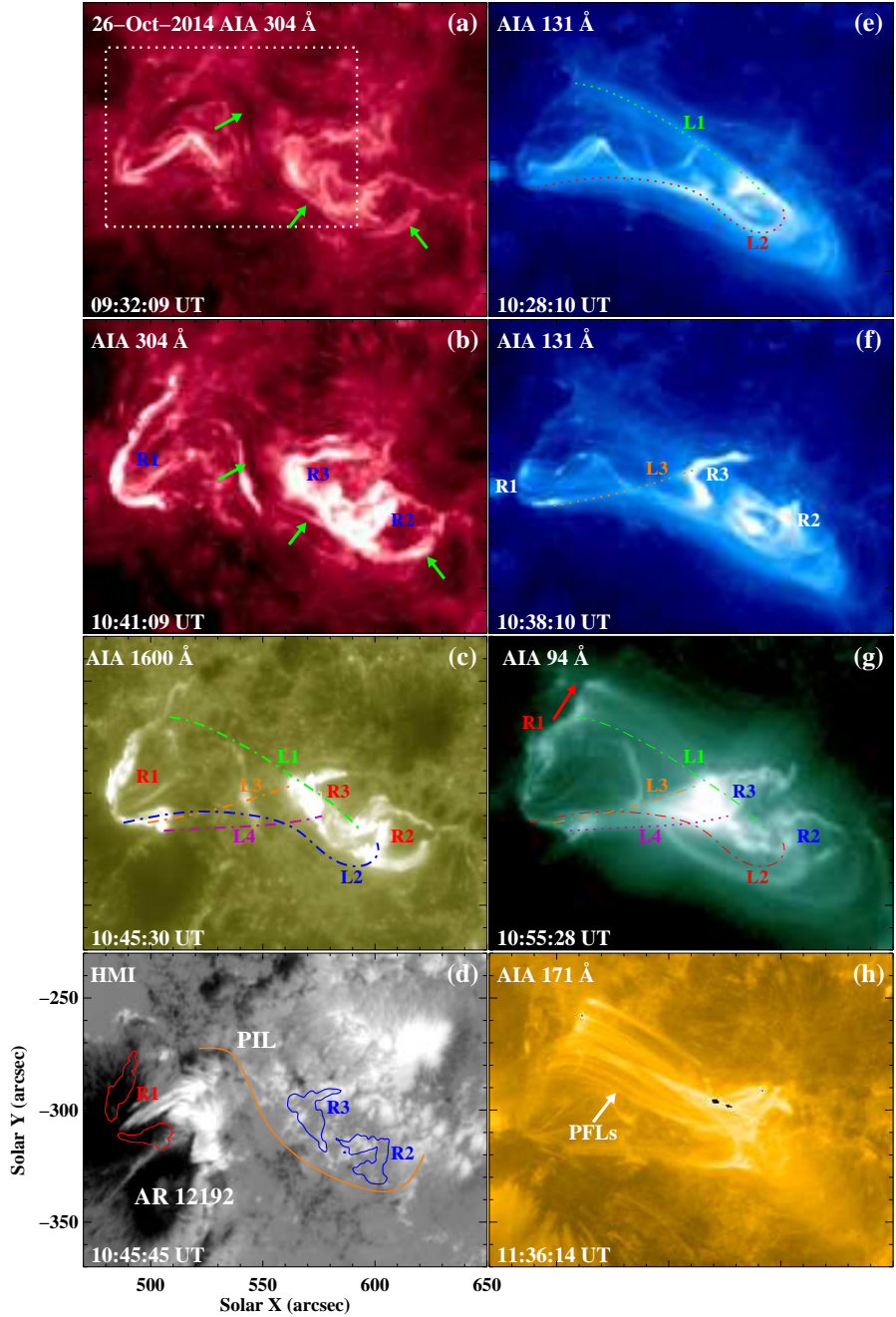}
\caption{\emph{SDO}/AIA multi-wavelengths (E)UV images and
\emph{SDO}/HMI LOS magnetogram showing the appearance of the
X2.0-class flare in AR 12192 on 2014 October 26. The green arrows in
panels (a)-(b) point to the non-eruptive filament along the PIL. The
white dotted rectangle in panel (a) denotes the FOV of Figures
8(a)-(c). R1-R3 are three flare ribbons in the central region of the
AR, and their brightness contours in 1600 {\AA} are shown in panel
(d). The orange curve in panel (d) indicates the average orientation
of the PIL of the AR. Dotted curves L1-L4 in panels (e)-(g) outline
the four loop bundles identified in 131 and 94 {\AA}, and their
duplicates are drawn by dash-dotted lines in panels (c) and (g). The
red arrow in panel (g) denotes the elongation motion of ribbon R1.
White arrow in panel (h) points to the post-flare loops in 171
{\AA}. The animation of this figure includes AIA 304, 171, 94, and
131 {\AA} images from 10:20 UT to 12:00 UT. The video duration is 25
s. \label{fig5}}
\end{figure}
\clearpage

\begin{figure}
\centering
\includegraphics
[bb=29 120 574 656,clip,angle=0,scale=0.8]{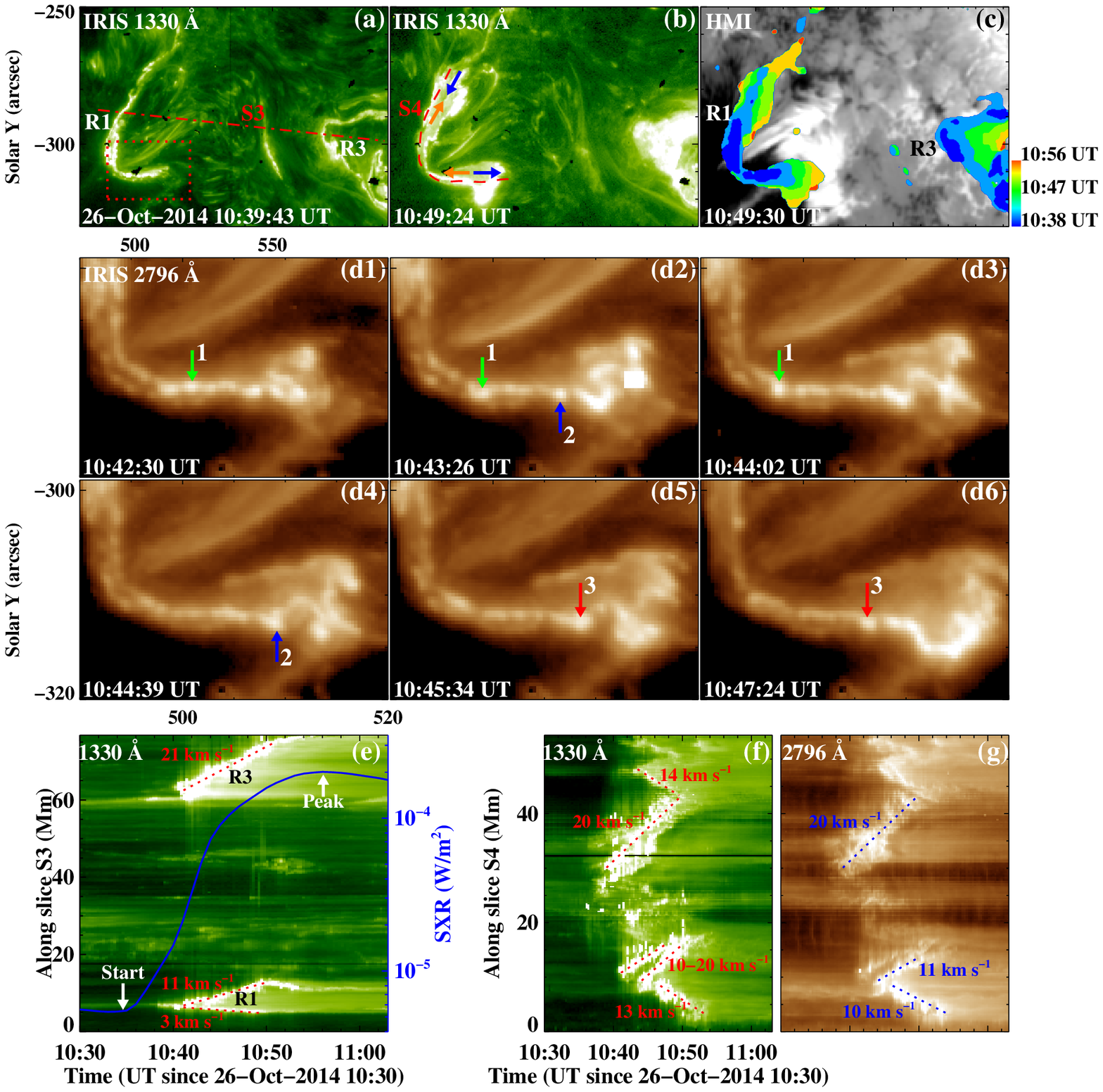}
\caption{Dynamic evolution of ribbons R1 and R3 from \emph{IRIS}
high-resolution observations. Panels (a)-(b): \emph{IRIS} 1330 {\AA}
images showing flare ribbons R1 and R3. Red dash-dotted line ``S3"
in panel (a) and dashed curve ``S4" in panel (b) respectively denote
the locations used to obtain the stack plots shown in panels (e) and
(f)-(g). Red dotted rectangle in panel (a) denotes the FOV of panels
(d1)-(d6). Orange and blue arrows in panel (b) represent the
bi-directional slippage of ribbon R1. Panel (c): \emph{SDO}/HMI LOS
magnetogram with an overlay of the evolving positions of ribbons R1
and R3. The color indicates the time of the ribbon brightness at
10:38-10:56 UT in 1330 {\AA} SJIs. Panels (d1)-(d6): time series of
\emph{IRIS} 2796 {\AA} images showing the slippage of traced bright
knots (``1"$-$``3") within ribbon R1. Bright knots ``1" and ``3"
slipped towards the east and knot ``2" slipped in the opposite
direction. Panel (e): 1330 {\AA} stack plot along slice ``S3" (panel
(a)) displaying the motions of R1 and R3 perpendicular to the PIL.
Blue curve is the GOES SXR 1-8 {\AA} flux of the flare. Panels
(f)-(g): 1330 and 2796 {\AA} stack plots along slice ``S4" (panel
(b)) showing the bi-directional slippage along ribbon R1. The
animation of this figure includes \emph{IRIS} 1330 and 2796 {\AA}
images from 10:00 UT to 11:15 UT. The video duration is 24 s.
\label{fig6}}
\end{figure}
\clearpage

\begin{figure}
\centering
\includegraphics
[bb=14 267 548 549,clip,angle=0,scale=0.8]{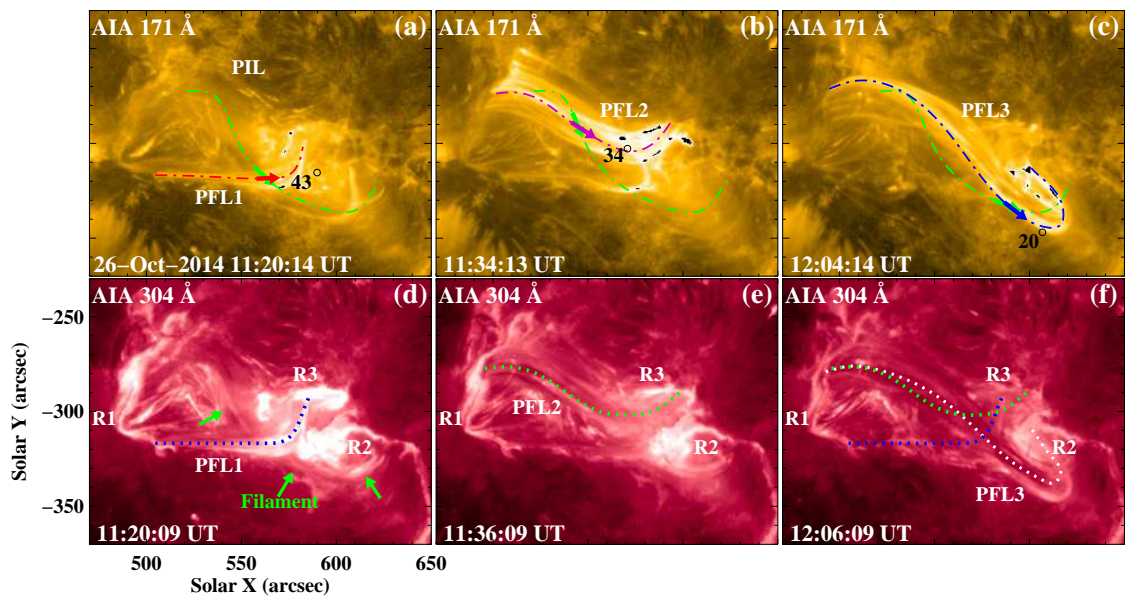} \caption{Time
series of 171 and 304 {\AA} images showing the sheared PFLs. Green
dash-dotted line is the duplication of the PIL (the orange curve in
Figure 7(d)). PFL1-PFL3 are three sets of PFLs, appearing
successively during the gradual phase of the flare. Three examples
of PFLs (red, purple and blue dash-dotted curves in panels (a)-(c))
are shown to estimate their inclination angles with respect to the
PIL. Green arrows in panel (d) point to the non-eruptive filament
underlying the PFLs. Blue, green and white dotted curves in panels
(d)-(f) represent the connectivities of dark arcades between ribbons
R1-R3. \label{fig5}}
\end{figure}
\clearpage

\begin{figure}
\centering
\includegraphics
[bb=107 262 471 550,clip,angle=0,scale=0.9]{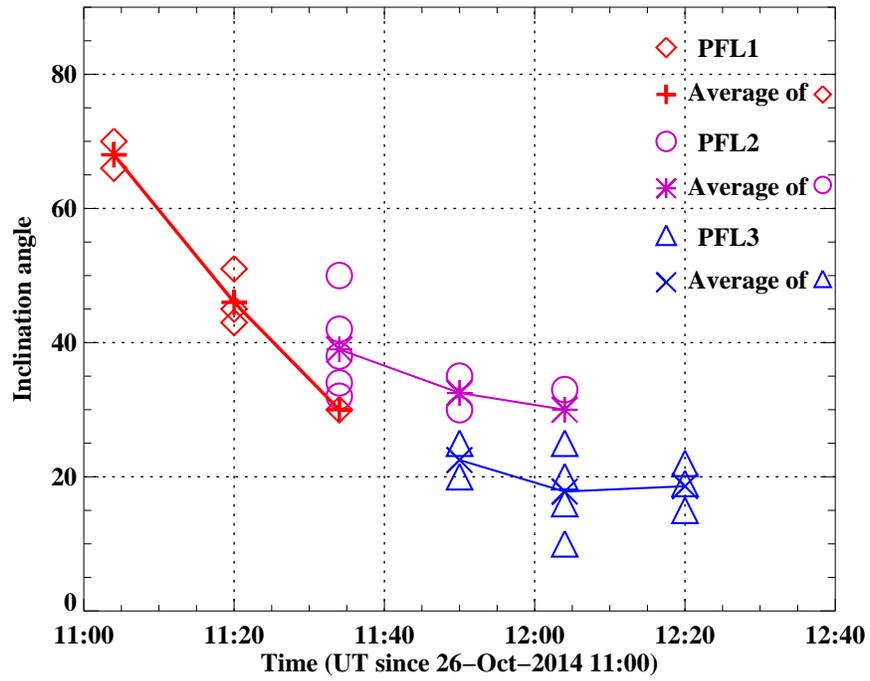}
\caption{Inclination angles of the observed PFLs with respect to the
PIL during the development of the flare. Red symbols denote the
inclination angles of the PFL1 and the average values. The purple
symbols are for the PFL2, and the blue marks for the PFL3.
\label{fig5}}
\end{figure}
\clearpage

\begin{figure}
\centering
\includegraphics
[bb=112 452 469 761,clip,angle=0,scale=1.0]{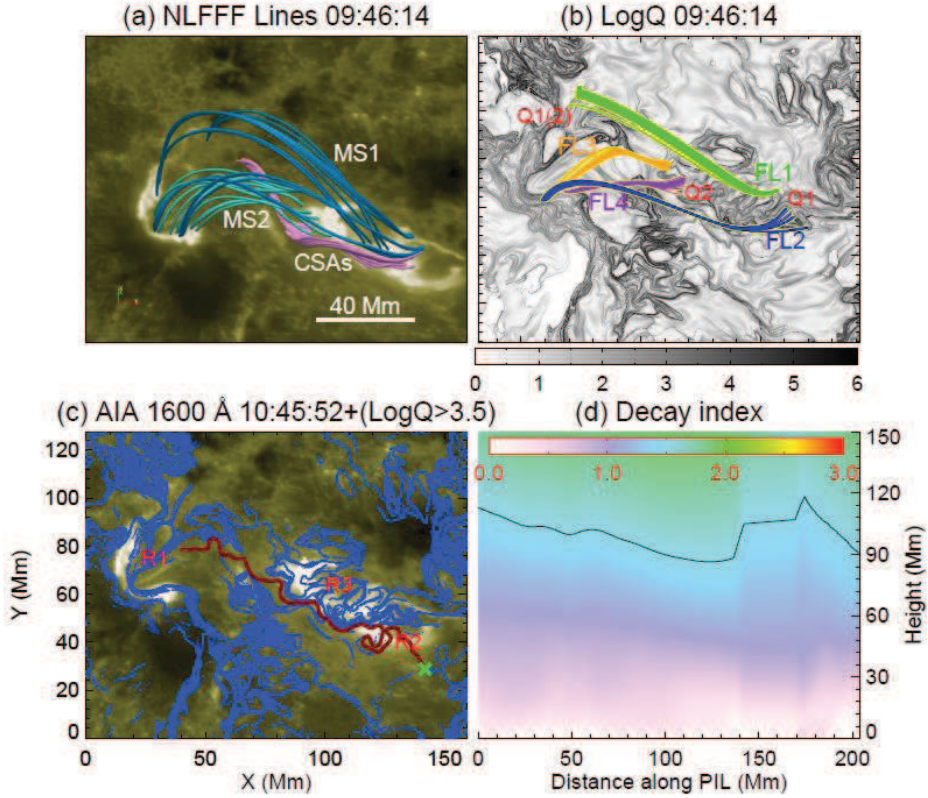}
\caption{Magnetic field structure, QSL and decay index distribution.
Panel (a): 3D structure of the field lines of the NLFFF showing the
CSAs (pink lines) and two sets of overlying magnetic systems (MS1
and MS2, blue and cyan lines). Panel (b): selected four strands of
field lines (FL1-FL4) plotted over the distribution of the squashing
factor Q in the z=0 plane. FL1-FL2 delineate a QSL structure Q1 and
FL3-FL4 outline another QSL structure Q2. Panel (c): overlay between
the flare ribbons (underlying 1600 {\AA} image) and the QSL map
(blue curves). All regions with Q$_{thresh}$$>$10$^{3.5}$ are shown
in this panel. The PIL is extracted from the bottom boundary of the
extrapolated NLFFF and marked by the red curve. The ``$\times$"
symbol shows the starting point of the PIL. Panels (b)-(c) have the
same FOV. Panel (d): distribution of the decay index n above the PIL
(red curve in panel (c)) prior to the flare onset. The black line
marks the positions where decay index n reaches the critical value
of 1.5. \label{fig7}}
\end{figure}
\clearpage

\begin{figure}
\centering
\includegraphics
[bb=103 181 463 633,clip,angle=0,scale=0.85]{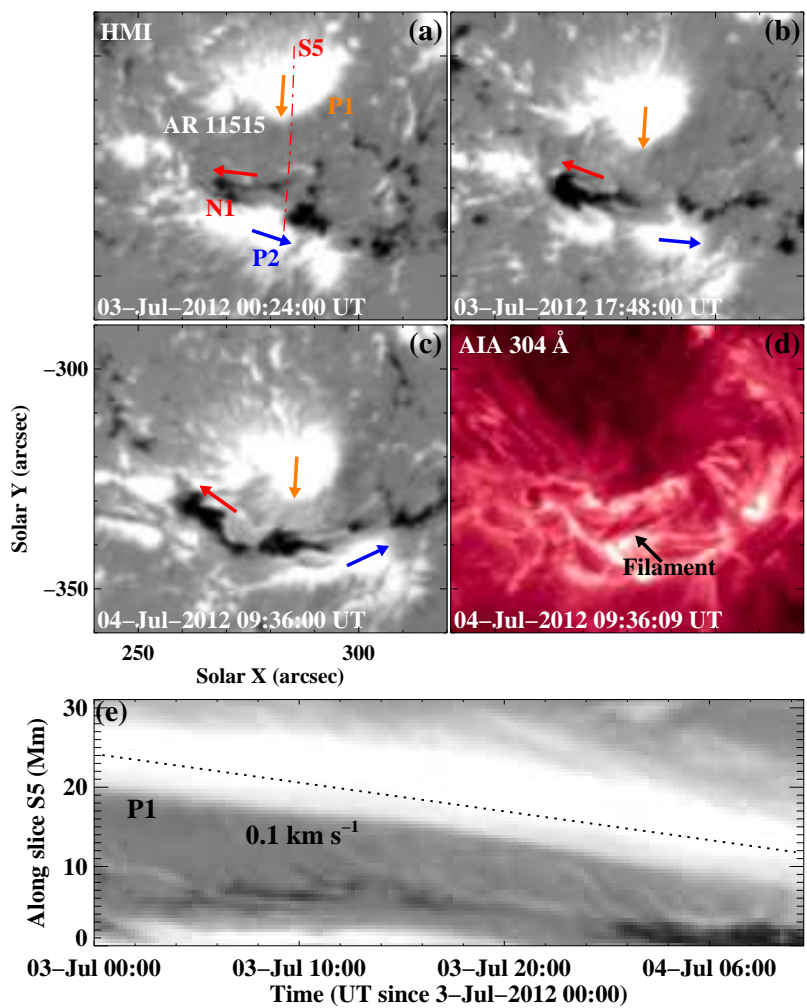}
\caption{Magnetic field evolution of AR 11515 prior to the
M5.3-class flare on 2012 July 04. Panels (a)-(c): \emph{SDO}/HMI LOS
magnetograms (saturating at $\pm$1000 G) showing the photospheric
evolution at the flaring region. Orange arrows represent the
converging motion of the positive-polarity sunspot P1. Red and blue
arrows denote the shearing motions of N1 and P2 in opposite
directions. The slice ``S5" is drawn as red dash-dotted line in
panel (a), for which the time-distance plot is presented in panel
(e). Panel (d): \emph{SDO}/AIA 304 {\AA} image showing the location
of the filament before the eruption. Panel (e): time-distance plot
along the slice ``S5" in the LOS magnetograms displaying the
southward convergence of P1. \label{fig8}}
\end{figure}
\clearpage

\begin{figure}
\centering
\includegraphics
[bb=15 171 549 644,clip,angle=0,scale=0.8]{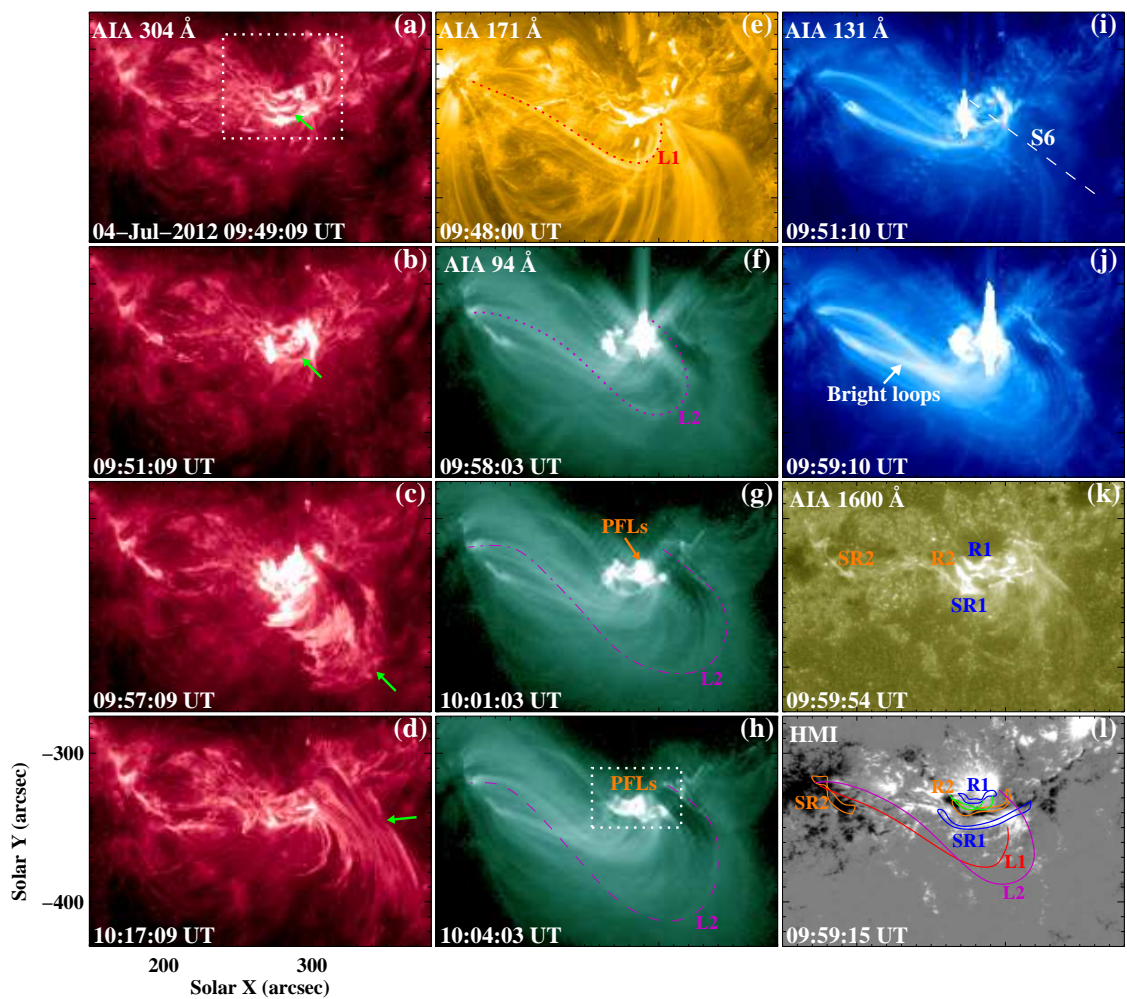}
\caption{\emph{SDO}/AIA multi-wavelengths (E)UV images and
\emph{SDO}/HMI LOS magnetogram showing the evolution of the
M5.3-class flare on 2012 July 04. Green arrows in panels (a)-(d)
point to the erupting filament involved in the event. White
rectangles in panels (a) and (h) respectively denote the FOVs of
Figures 12(a)-(d) and 14(a)-(c). Red and purple curves in panels
(e)-(h) outline two large-scale loop bundles L1 and L2 overlying the
flaring region. PFLs in panels (g)-(h) are post-flare loops
underlying the eruptive filament in 94 {\AA}. Dashed line ``S6"
(panel (i)) represents the location used to obtain the stack plot
shown in Figures 14(d)-(e). R1 and R2 are two main ribbons, and SR1
and SR2 are two secondary ribbons. The pre-eruption filament (green
lines), four flare ribbons (blue and orange contours) and loop
bundles (red and purple lines) are plotted over the LOS magnetogram
in panel (l). The animation of this figure includes AIA 171, 304, 94
and 131 {\AA} images from 09:30 UT to 11:00 UT. The video duration
is 22 s. \label{fig9}}
\end{figure}
\clearpage

\begin{figure}
\centering
\includegraphics
[bb=15 249 560 535,clip,angle=0,scale=0.75]{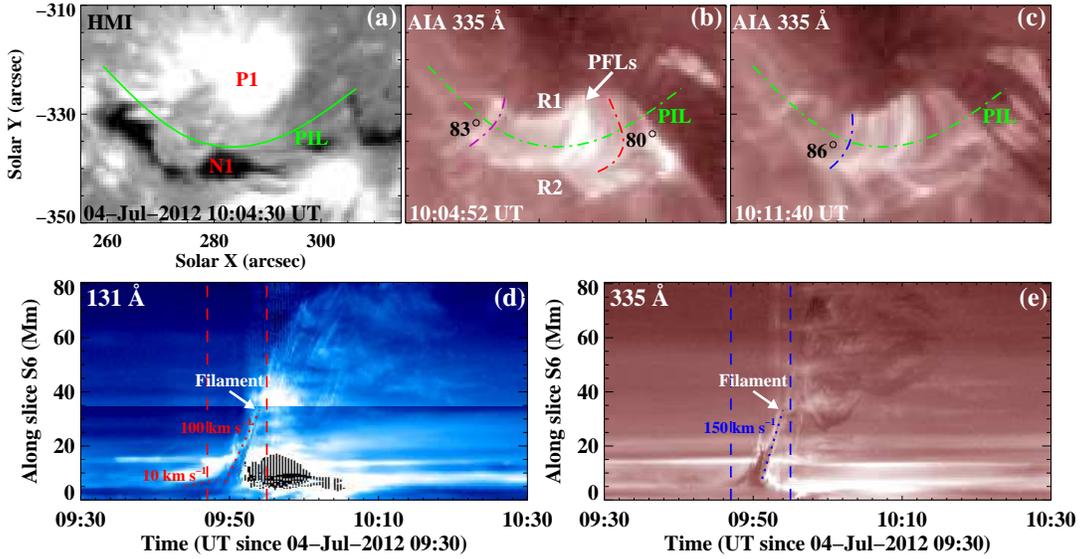}
\caption{Appearance of post-flare loops (PFLs) and kinematic
evolution of the erupting filament. Panels (a)-(c): \emph{SDO}/HMI
LOS magnetogram and \emph{SDO}/AIA 335 {\AA} images showing the
average orientation of the PIL (green curves) of the flaring region
and the approximately potential PFLs. P1 and N1 in panel (a) are two
magnetic structures where two main ribbons R1 and R2 are located.
Purple, red and blue curves in panels (b)-(c) outline three PFLs and
their inclination angles with respect to PIL are also shown. Panels
(d)-(e): time-distance plots along slice ``S6" (Figure 13(i)) in 131
and 335 {\AA} displaying the kinematic evolution of the erupting
filament. Two dashed lines respectively denote the start and peak
times of the flare. \label{fig10}}
\end{figure}
\clearpage

\begin{figure}
\centering
\includegraphics
[bb=15 196 575 591,clip,angle=0,scale=0.75]{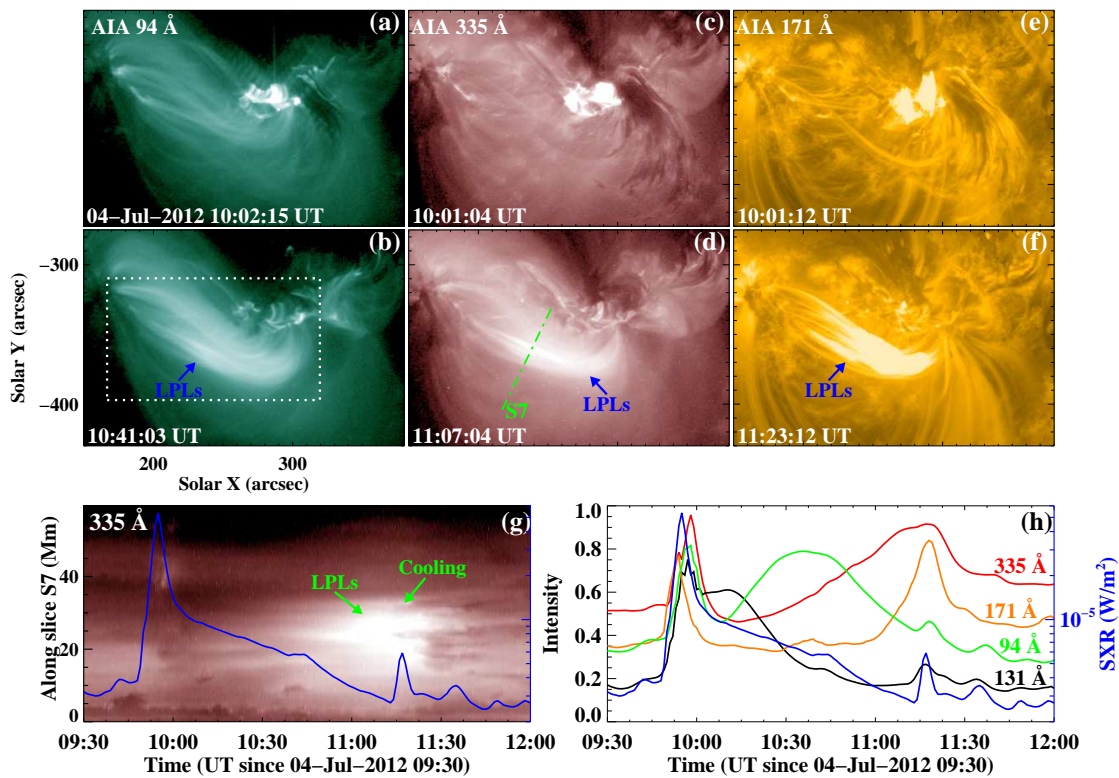} \caption{Late
phase of the M5.3-class flare. Panels (a)-(f): \emph{SDO}/AIA 94,
335 and 171 {\AA} images showing the brightening late-phase loops
(LPLs) in different wavelengths. The white box in panel (b) is used
to obtain the emission variations as shown in panel (h). ``S7" in
panel (d) is used to obtain the time-distance plot shown in panel
(g). Panels (g)-(h): time-distance plot along slice ``S7" in 335
{\AA} and the EUV emission variations within the white box (panel
(a)) showing the late phase of the flare. Blue curves are the GOES
SXR 1-8 {\AA} flux variation of the flare. \label{fig10}}
\end{figure}
\clearpage

\begin{figure}
\centering
\includegraphics
[bb=115 497 498 761,clip,angle=0,scale=0.9]{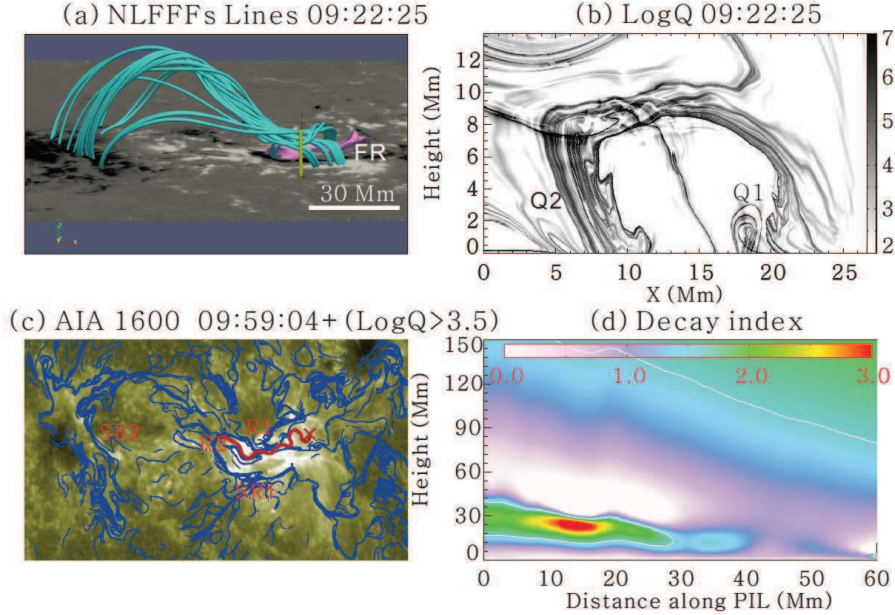}
\caption{Magnetic field structure and distributions of QSL and decay
index. Panel (a): side view of the 3D magnetic topology of the
pre-explosion AR at about 09:22 UT on 2012 July 04 resulting from
the NLFFF extrapolation. Pink lines are the flux rope (FR) field
lines, and cyan lines correspond to the overlying strapping fields.
The yellow bar marks the position and extent of the vertical plane
where panel (b) is obtained. Panel (b): Q distribution in a vertical
plane across the pre-eruptive FR. Q1 is the FR-related QSL and Q2 is
the dome-shaped QSL encircling Q1. Panel (c): overlay between the
flare ribbons (underlying 1600 {\AA} image) and the QSL map (blue
curves). All regions with Q$_{thresh}$$>$10$^{3.5}$ are shown in
this panel. The PIL between two main ribbons R1 and R2 is extracted
from the bottom boundary of the extrapolated NLFFFs and marked by
the red curve. The ``$\times$" symbol shows the starting point of
the PIL. Panel (d): distribution of the decay index above the PIL
(red curve in panel (c)) prior to the flare onset. The white lines
mark the positions where decay index n reaches the critical value of
1.5. \label{fig11}}
\end{figure}
\clearpage

\begin{figure}
\centering
\includegraphics
[bb=66 12 545 386,clip,angle=0,scale=0.8]{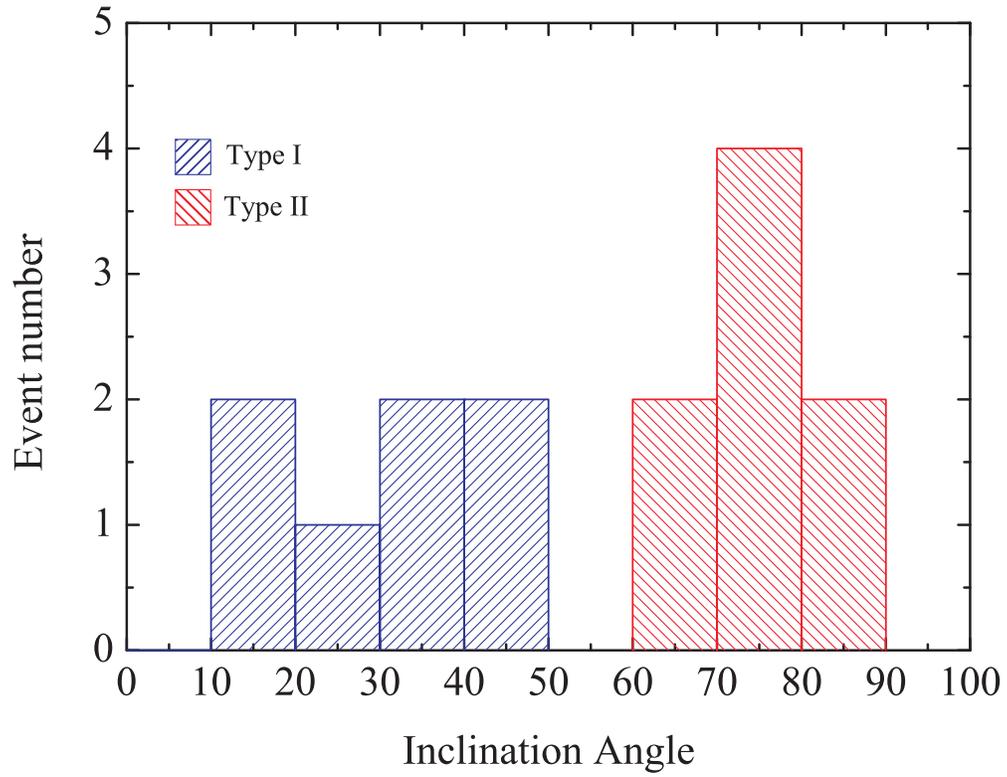}
\caption{Histogram of measured inclination angle of PFLs for the 15
confined flares. Blue/red represents ``type I"/``type II" flares.
The bin size in the histogram is $10^{\circ}$. \label{fig15}}
\end{figure}
\clearpage

\end{document}